\documentclass[preprint,12pt]{elsarticle}

\usepackage{graphicx}
\usepackage{booktabs}
\usepackage{multirow}
\usepackage{amsmath}
\usepackage{listings}
\usepackage{hyperref}
\usepackage{bookmark} 
\usepackage{geometry}
\usepackage[linesnumbered,ruled,vlined]{algorithm2e} 
\usepackage{float}
\usepackage{setspace}
\usepackage[T1]{fontenc}
\usepackage{lmodern}
\usepackage{textcomp}
\usepackage{placeins}
\usepackage[table]{xcolor}
\usepackage{adjustbox}
\usepackage{chngcntr}
\counterwithout{table}{section}
\usepackage{makecell}
\usepackage{ragged2e} 

\usepackage{tabularx}

\usepackage{lineno}
\usepackage{lmodern}
\usepackage{color}
\usepackage{listings}
\usepackage{tabularx}
\usepackage{enumerate}
\usepackage{enumitem}
\usepackage{subcaption} 
\usepackage[normalem]{ulem}

\usepackage{multirow}

\definecolor{Comments}{rgb}{0.00,0.50,0.00}
\definecolor{KeyWords}{rgb}{0.00,0.00,0.13}
\definecolor{Strings}{rgb}{0.60,0.00,0.00}

\lstdefinestyle{psql}
{
  tabsize=3,
  basicstyle=\small\upshape\ttfamily,
  language=SQL,
  morekeywords={PROVENANCE,BASERELATION,INFLUENCE,COPY,ON,TRANSPROV,TRANSSQL,TRANSXML,CONTRIBUTION,COMPLETE,TRANSITIVE,NONTRANSITIVE,EXPLAIN,SQLTEXT,GRAPH,IS,ANNOT,THIS,XSLT,MAPPROV,cxpath,OF,TRANSACTION,SERIALIZABLE,COMMITTED,INSERT,INTO,WITH,SCN,PROV,IMPORT,FOR,JSON,JSON_TABLE,XMLTABLE,DATABASE,SYSTEM,SWITCH,LOGFILE,MEMBER},
  extendedchars=false,
  keywordstyle=\color{blue},
  mathescape=true,
  escapechar=@,
  sensitive=true,
  stringstyle=\color{Strings},%
  string=[b]'
}

\lstdefinestyle{rsl}
{
tabsize=3,
basicstyle=\small\upshape\ttfamily,
language=C,
morekeywords={RULE,LET,CONDITION,RETURN,AND,FOR,INTO,REWRITE,MATCH,WHERE},
extendedchars=false,
keywordstyle=\color{blue},
mathescape=true,
escapechar=@,
sensitive=true
}

\lstdefinestyle{pseudocode}
{
  tabsize=3,
  basicstyle=\small,
  language=c,
  morekeywords={if,else,foreach,case,return,in,or},
  extendedchars=true,
  mathescape=true,
  literate={:=}{{$\gets$}}1 {<=}{{$\leq$}}1 {!=}{{$\neq$}}1 {append}{{$\listconcat$}}1 {calP}{{$\cal P$}}{2},
  keywordstyle=\color{blue},
  escapechar=&,
  numbers=left,
  numberstyle={\color{green}\small\bf}, 
  stepnumber=1, 
  numbersep=5pt,
}

\lstdefinestyle{xmlstyle}
{
  tabsize=3,
  basicstyle=\small,
  language=xml,
  extendedchars=true,
  mathescape=true,
  escapechar=£,
  tagstyle={\color{blue}},
  usekeywordsintag=true,
  morekeywords={alias,name,id},
  keywordstyle={\color{red}}
}

\lstdefinelanguage{json}{
    basicstyle=\footnotesize\ttfamily,
    numbers=left,
    numberstyle=\scriptsize,
    stepnumber=1,
    numbersep=8pt,
    stringstyle=\color{Strings},%
    showstringspaces=false,
    breaklines=true,
    frame=lines,
    string=[b]"
}

\newcommand{\NoCarver}{\texttt{ANOC}}
\newcommand{\radar}{\texttt{RADAR}}

\makeatletter
\def\ps@pprintTitle{%
   \let\@oddhead\@empty
   \let\@evenhead\@empty
   \def\@oddfoot{\reset@font\hfil\thepage\hfil\textit{Preprint}}%
   \let\@evenfoot\@oddfoot}
\makeatother

\begin{document}

\begin{frontmatter}

\title{RADAR: Exposing Unlogged NoSQL Operations}

\author{Mahfuzul I. Nissan}
\ead{minissan@uno.edu}
    
\author{James Wagner}
\ead{jwagner4@uno.edu}

\cortext[corresponding]{Corresponding author}

\address{University of New Orleans, New Orleans, USA}

\begin{abstract}



The widespread adoption of NoSQL databases has made digital forensics increasingly difficult as storage formats are diverse and often opaque, and audit logs cannot be assumed trustworthy when privileged insiders, such as DevOps or administrators, can disable, suppress, or manipulate logging to conceal activity. We present \radar{} (Record \& Artifact Detection, Alignment \& Reporting), a log-adversary-aware framework that derives forensic ground truth by cross-referencing low-level storage artifacts against high-level application logs. \radar{} analyzes artifacts reconstructed by the Automated NoSQL Carver (\NoCarver{}), which infers layouts and carves records directly from raw disk bytes, bypassing database APIs and the management system entirely, thereby treating physical storage as the independent evidence source. \radar{} then reconciles carved artifacts with the audit log to identify delta artifacts such as unlogged insertions, silent deletions, and field-level updates that exist on disk but are absent from the logical history. We evaluate \radar{} across ten NoSQL engines, including BerkeleyDB, LMDB, MDBX, etcd, ZODB, Durus, LiteDB, Realm, RavenDB, and NitriteDB, spanning key-value and document stores and multiple storage designs, e.g., copy-on-write/MVCC, B/B\texttt{+}-tree, and append-only. Under log-evasion scenarios, such as log suppression and post-maintenance attacks, including cases where historical bytes are pruned, \radar{} consistently exposes unattributed operations while sustaining 31.7–397 MB/min processing throughput, demonstrating the feasibility of log-independent, trustworthy NoSQL forensics.

\end{abstract}


\end{frontmatter}
\onehalfspacing


\section{Introduction}
\label{sec:introduction}
Database Management Systems (DBMS) are foundational for storing and processing sensitive data. Accordingly, extensive effort has been devoted to securing these systems through access control and audit logging. However, once a user gains elevated privileges, whether legitimately or by exploiting a vulnerability, these security measures can be circumvented. An attacker can manipulate data and then alter or disable logging to conceal the activity. Forensic analysis therefore requires reliable methods to detect breaches and collect evidence about unauthorized activity when logs cannot be fully trusted.

A more trustworthy source of evidence lies in the low-level storage itself. Any operation, whether insertion, update, or deletion, leaves traces in the physical storage layer. While attackers can modify a log file, it is exceptionally difficult to tamper with byte-level database structures on disk or in RAM without causing inconsistencies or corruption. This makes direct storage analysis a powerful method for providing independent, storage-level evidence of database activity.

While prior research has addressed storage analysis for relational databases, the challenge is particularly acute for NoSQL systems. The NoSQL ecosystem encompasses a wide array of architectures, each with proprietary, non-standardized, and often undocumented storage formats. A malicious actor with sufficient privileges can exploit this complexity. By modifying startup scripts or deployment configurations, such an actor can suspend logging, perform unlogged operations, and then restore the logging configuration, leaving a gap in the audit log that is difficult to detect through conventional monitoring.

To address these challenges, we introduce \radar{}, a unified framework designed for forensic carving and log verification in NoSQL databases. This framework systematically analyzes low-level storage artifacts and cross-verifies them with application audit logs. \radar{} utilizes our Automated NoSQL Carver \cite{nissan2025anoc} (ANOC) to automatically infer storage layouts and carve records from disk and memory. By comparing these carved artifacts to audit logs, \radar{} identifies inconsistencies, such as unlogged modifications, silent deletions, or field-level tampering. The primary contributions of this work are:

\begin{itemize}
    \item We introduce a log-adversary-aware framework (\radar{}) that unifies storage carving with audit log verification to detect unattributed operations in NoSQL databases (Section \ref{sec:framework}).

    \item We propose two detection methodologies tailored to different storage architectures: single-snapshot analysis for versioned systems and comparative analysis for in-place modification systems (Section \ref{sec:methodology}).
    
    \item We formalize algorithms for identifying unattributed insertions, deletions, and updates through reconciliation of carved artifacts with audit logs (Section \ref{sec:methodology}).
    
    \item We evaluate \radar{} across ten NoSQL databases (both key-value and document stores) under diverse tampering scenarios including post-maintenance attacks and log suppression (Section \ref{sec:experiments}).
\end{itemize}
\section{Related Work}
\label{sec:relatedWork}

The forensic analysis of database systems spans two long-running threads: recovering \emph{what} is physically present in storage and auditing \emph{how} it purportedly got there. We review storage-level reconstruction for relational and NoSQL engines, then survey log integrity and provenance work that motivates cross-checking logs against independent ground truth.

\subsection{Forensic Recovery from Database Storage}
\label{subsec:storage_recovery}

Relational DBMSes provide well-documented page layouts and centralized catalogs, such as MySQL’s INFORMATION\_SCHEMA or PostgreSQL’s pg catalog, which investigators can leverage to parse pages and row directories \cite{mysql_information_schema, hellerstein2007architecture}. Early forensic work focused on exploiting this regularity to reconstruct damaged or deleted records from individual engines such as SQL Server or InnoDB \cite{9328241, fruhwirt2012innodb}. Building on these database-specific methods, Wagner et al. introduced internal structure carving as a more systematic approach, later extended into generalized tools like DBCarver \cite{wagner2015database, wagner2017database, wagner2019db3f, wagner2016database}.

Beyond page carving, additional efforts expanded the forensic toolkit. Frühwirt et al. reconstructed data manipulation queries by analyzing InnoDB redo logs, demonstrating the value of WAL artifacts in relational investigations \cite{fruhwirt2012innodb}. File carving research also influenced the field, with Garfinkel formalizing fast validation of carved objects and Poisel and Tjoa providing a comprehensive survey of carving techniques \cite{garfinkel2007carving, poisel2013file}. More recently, Lenard et al. proposed Sysgen, a system state corpus generator for synthetic testing of forensic methods, which helped establish controlled baselines for evaluation \cite{lenard2020sysgen}.

NoSQL engines disrupt these assumptions. Storage designs vary widely: Berkeley DB, LMDB, and WiredTiger rely on page-based B/B$^+$-tree structures, whereas ZODB uses serialized object streams \cite{wiredtiger_btrees, ZODB, candel2022unified}. The absence of unified catalogs makes reconstruction more difficult than in relational systems. Prior work therefore focused on individual databases, such as case studies of MongoDB or Redis, each requiring its own specialized tools \cite{yoon2016forensic, yoon2018method, sung2019less, xu2014forensic}. To address this heterogeneity, our earlier work introduced ANOC, which dynamically infers structural parameters and reconstructs records directly from raw bytes \cite{nissan2025anoc}. While ANOC establishes the \emph{physical presence} of data, it does not determine whether that data is legitimate given the declared history of operations.

\subsection{Log Analysis, Integrity, and Tamper Detection}
\label{subsec:log_analysis}

Audit logs remain the most common source for reconstructing timelines of user actions. Yet their reliability is limited in adversarial settings, since privileged attackers can suppress or selectively edit entries \cite{PS08, snodgrass2004tamper}. This challenge is particularly acute in the NoSQL ecosystem. Unlike many relational databases that offer robust, built-in auditing frameworks, NoSQL systems often lack native logging capabilities or relegate them to enterprise-only features. This forces the responsibility of generating a verifiable audit trail onto application developers, resulting in heterogeneous, application-level logs that are more susceptible to bypass and manipulation.

Researchers have long recognized this risk and proposed numerous solutions for log integrity. Schneier and Kelsey introduced seminal work on secure, forward-secure audit logs to resist retroactive modification \cite{SchneierKelsey1999}. Crosby and Wallach later developed efficient tamper-evident data structures using Merkle-tree chaining \cite{crosby2009efficient}. In parallel, standards bodies issued guidance, with NIST’s SP 800-92 emphasizing centralized log management and integrity verification \cite{Kent2007NIST}. Hardware-assisted designs provide another direction, using trusted execution environments or dedicated hardware to create tamper-proof audit trails \cite{paccagnella2020custos, ahmad2022hardlog}, though these depend on specialized deployments.

A complementary line of research turns to provenance. Provenance-based intrusion detection systems capture causal relationships across system-level events. CamFlow demonstrated practical whole-system provenance capture, and HOLMES used graph correlation to detect advanced persistent threats \cite{pasquier2017camflow, milajerdi2019holmes}. Such systems reveal inconsistencies at the OS level, though they stop short of validating record-level database contents. This gap motivates approaches that compare audit logs against independent physical evidence.

\subsection{A Unified Forensic Approach}
\label{subsec:unified_approach}

Although storage artifacts and audit logs have traditionally been studied in isolation, there is growing recognition that combining them strengthens tamper detection. Wagner et al. showed that page-level carving could expose file tampering in relational systems \cite{wagner2018detecting}. More recent work in database memory forensics reconstructed query semantics or cache access patterns to serve as a parallel verification stream \cite{NISSAN2023301503, nissan2022forensic, WAGNER2023301567}. Building on these insights, our \radar{} framework is the first to systematically align generalized, carved NoSQL artifacts with normalized application audit logs to identify \emph{unattributed operations}. By reconciling two independent evidence streams, \radar{} detects anomalies such as unlogged deletions, silent updates, or rogue insertions that would remain invisible to either storage carving or log auditing alone.
\section{Threat Model}
\label{sec:threat-model}

\begin{figure*}[!t]
\centering
\includegraphics[width=1\textwidth]{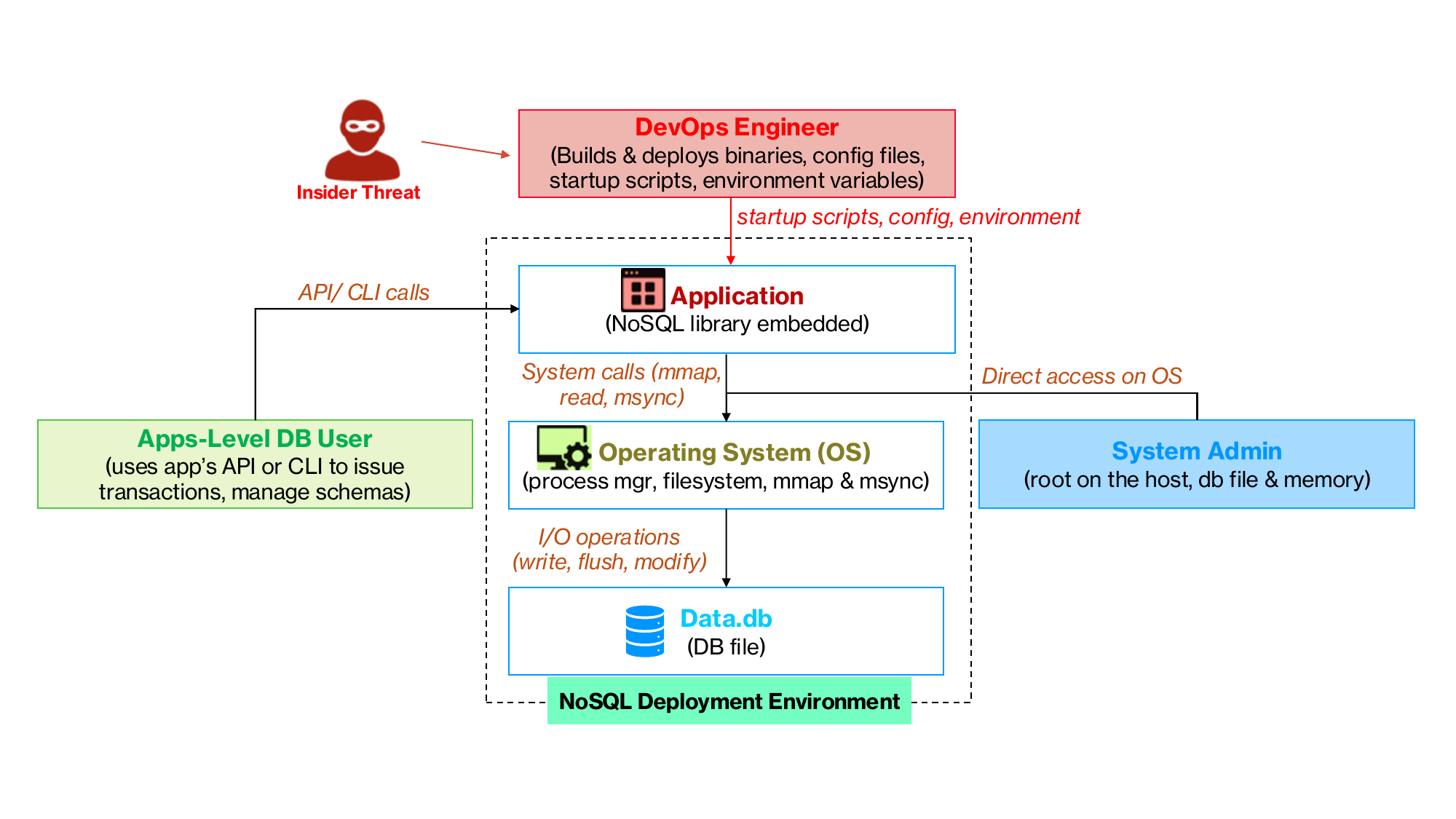}
\vspace{-4mm}
\caption{Threat model}
\vspace{-4mm}
\label{fig:threat_model}
\end{figure*}

In a typical NoSQL deployment, multiple operational roles interact with the system at different privilege layers, as shown in Figure~\ref{fig:threat_model}. We distinguish three such roles according to their operational scope and privilege boundaries: the Application-Level Database User (ALDU), the DevOps Engineer (DO), and the System Administrator (SA). The ALDU is a legitimate operator who interacts with the database exclusively through the application’s exposed interfaces, such as an API or command-line client. This role may issue queries and perform transactions as permitted by application logic, with all operations mediated by application-enforced access control and auditing. The ALDU cannot modify deployment artifacts, directly access storage files, or alter operating system configurations, and is considered a non-malicious actor in this work. The DO, in contrast, is responsible for building and deploying the NoSQL application, including its binaries, configuration files, and startup scripts. By controlling runtime instantiation and operational parameters, a malicious DO can introduce deployment-layer tampering that bypasses application-level auditing. However, the DO does not possess root or kernel-level privileges, which remain under the control of the SA. The SA operates the host OS with full root privileges, manages process control, filesystem permissions, and storage devices, and is responsible for installing OS-level integrity monitoring and snapshot mechanisms. While the SA can access database storage files directly, they do not control the build or deployment process. In our model, the SA is a trusted role that maintains host-level monitoring independent of the application layer.


Given these privilege boundaries, the adversary in this model is the DO. Their access to the deployment pipeline enables three primary forms of abuse:

\begin{enumerate}
  \item \textbf{Configuration abuse:} Manipulating configuration to disable or degrade audit logging, performing unauthorized operations, and then restoring the original settings.
  \item \textbf{Backdoored deployment:} Deploying a backdoored application binary that suppresses or falsifies audit events while performing hidden writes.
  \item \textbf{Out-of-band modification:} Executing an out-of-band script or tool to modify the database file directly, bypassing the official application process.
\end{enumerate}

Each pathway evades application-layer controls yet inevitably produces observable side effects at the operating system boundary. Examples include file writes from unexpected processes, binary integrity mismatches, and anomalous drift between inbound requests and bytes written. Crucially, because the DO lacks SA-level privileges, they cannot suppress or remove these OS-level traces without escalating beyond their assigned role.

Under this adversary, application and NoSQL-layer audit logs are not a reliable source of truth. Deployment-time configuration, controlled by the DO, can suspend, redirect, or downsample logging (e.g., changes to container logging drivers or service units), producing routine-looking rollouts with gaps in coverage. Many engines generate plaintext or JSON logs to ordinary files or external sinks without cryptographic binding to the executing binary or to trusted time, enabling truncation or selective edits by a privileged deployer. Moreover, environment and loader manipulation (e.g., preloaded libraries or wrapper launchers) can interpose on logging paths without modifying the application itself. Consequently, logs may be absent, selectively suppressed, or rewritten. In our model, they serve only as corroborating signals, while evidentiary completeness is derived from storage-level artifacts recovered by \NoCarver{}.

To counter these threats, we assume the SA can trigger forensic snapshot acquisition outside the application using host-controlled telemetry and change-control policy. In practice, the SA may rely on OS-level auditing or provenance signals (e.g., Linux \texttt{auditd} / \texttt{fanotify} / eBPF; Windows SACL / ETW / Sysmon) to attribute database-file modifications to a writing process (e.g., PID, UID / SID, executable path, and file path or inode). When available, executable integrity can be validated using host mechanisms such as Linux IMA or Windows code-integrity controls (e.g., WDAC), but \radar{} does not require per-write hashing guarantees. The operating system also provides native snapshot mechanisms (e.g., LVM / ZFS / Btrfs on Linux or Volume Shadow Copy on Windows). The SA encodes change-control rules and triggers a snapshot when policy indicates elevated forensic risk.

Under this model, writes from the approved database service under expected workload conditions are treated as \emph{normal} operations and may be logged without snapshot acquisition. Writes corresponding to authorized bulk changes under an approved change request, accompanied by a declared migration identifier, are treated as \emph{maintenance} operations and can trigger informational snapshots for documentation and post-facto auditing. Any write that violates provenance or change-control expectations (e.g., a non-service writer, audit suppression / toggling, integrity mismatches, unexpected service restarts outside a maintenance window, or unexplained drift anomalies) is treated as \emph{suspicious} and triggers immediate snapshot acquisition. Importantly, suspiciousness is defined by violations in provenance or control-plane policy rather than raw write volume alone, enabling higher-fidelity snapshot triggering without overwhelming the system with benign activity.

\begin{figure*}[!t]
\centering
\includegraphics[width=0.7\textwidth]{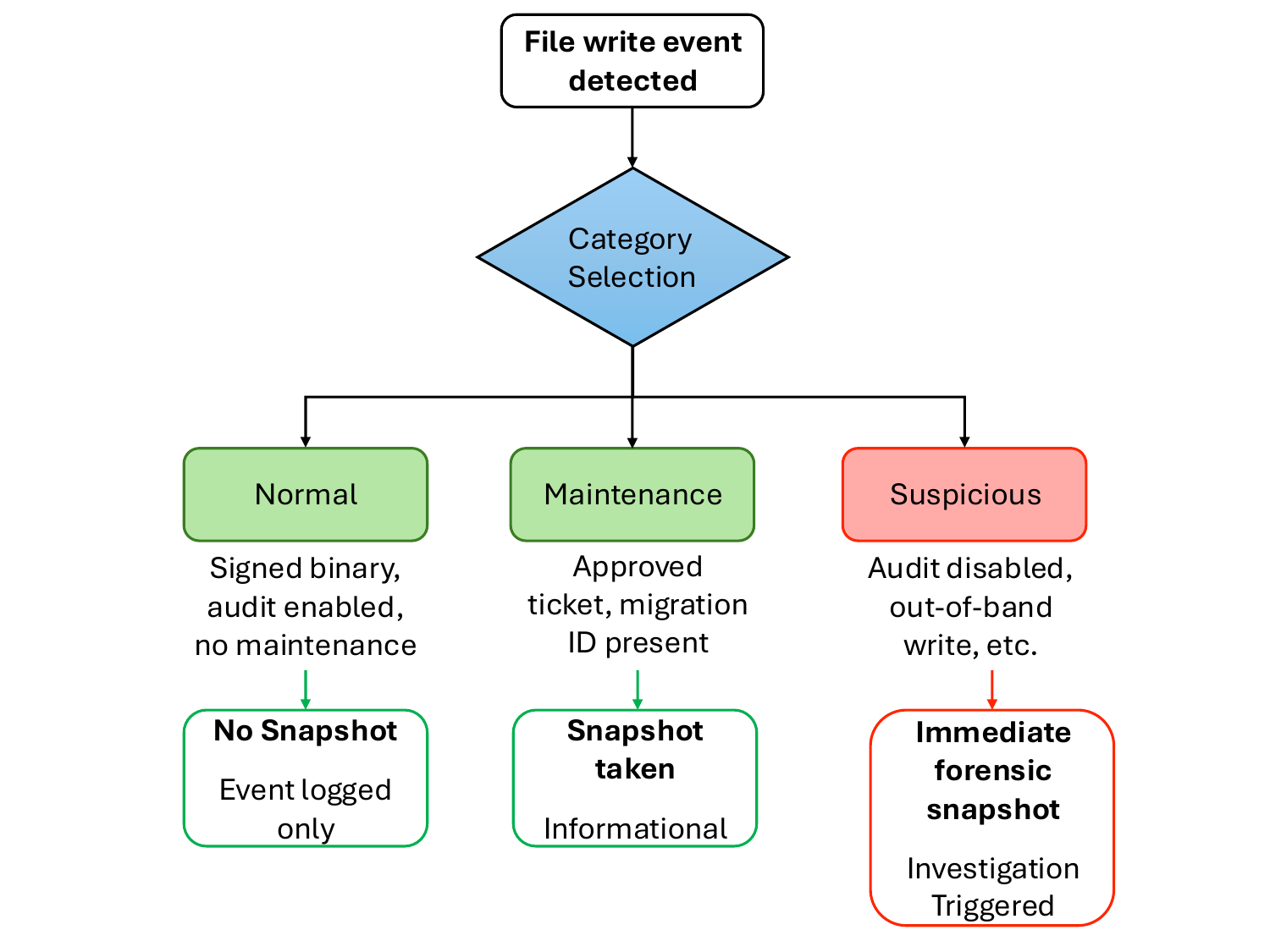}
\vspace{-4mm}
\caption{File-write event classification and the resulting snapshot decision.}
\vspace{-4mm}
\label{fig:file_write_event_detection}
\end{figure*}

Figure~\ref{fig:file_write_event_detection} and Table~\ref{table:snapshot_policy} summarize this classification and snapshot-trigger process. When the SA receives an OS audit event indicating a write to the database file (e.g., from \texttt{auditd} or Sysmon), it assigns the event to one of three categories (\emph{Normal}, \emph{Maintenance}, or \emph{Suspicious}) based on provenance and policy checks. This classification determines whether the event is only logged, results in an informational snapshot, or triggers immediate forensic snapshot acquisition followed by investigation.

\begin{table}[h!]
\centering
\small
\renewcommand{\arraystretch}{1.25} 
\setlength{\tabcolsep}{4pt}

\begin{adjustbox}{max width=\textwidth,center}
\begin{tabular}{|>{\raggedright\arraybackslash}p{2.2cm}|
                >{\raggedright\arraybackslash}p{6.4cm}|
                >{\raggedright\arraybackslash}p{3.0cm}|
                >{\raggedright\arraybackslash}p{4.0cm}|}
\hline
\rowcolor[HTML]{E6E6E6}
\multicolumn{1}{|>{\centering\arraybackslash}p{2.2cm}|}{\textbf{Category}} &
\multicolumn{1}{>{\centering\arraybackslash}p{6.4cm}|}{\textbf{Conditions}} &
\multicolumn{1}{>{\centering\arraybackslash}p{3.0cm}|}{\textbf{Snapshot Action}} &
\multicolumn{1}{>{\centering\arraybackslash}p{4.0cm}|}{\textbf{Purpose}} \\ \hline

Normal &
Approved, signed binary; audit enabled; workload within expected bounds; no maintenance window &
Log only &
Minimize false positives from routine operations \\ \hline

Maintenance &
Approved change ticket; signed binary; audit enabled; declared migration identifier &
Snapshot (informational) &
Document authorized changes for forensic completeness \\ \hline

Suspicious &
Any of: audit disabled/toggled; binary mismatch at start; non-service writer (out-of-band); unexpected stop/start or config change outside window; drift anomalies without ticket; writes into protected namespaces outside maintenance &
Immediate forensic snapshot &
Capture potential insider tampering for investigation \\ \hline

\end{tabular}
\end{adjustbox}
\vspace{-0.3cm}
\caption{Snapshot trigger policy}
\label{table:snapshot_policy}
\end{table}

\begin{table}[h!]
\centering
\small
\renewcommand{\arraystretch}{1.25} 
\setlength{\tabcolsep}{4pt}

\begin{adjustbox}{max width=\textwidth,center}
\begin{tabular}{|>{\raggedright\arraybackslash}p{1.6cm}|
                >{\raggedright\arraybackslash}p{5.8cm}|
                >{\raggedright\arraybackslash}p{4.2cm}|
                >{\raggedright\arraybackslash}p{4.0cm}|}
\hline
\rowcolor[HTML]{E6E6E6}
\multicolumn{1}{|>{\centering\arraybackslash}p{1.6cm}|}{\textbf{Platform}} &
\multicolumn{1}{>{\centering\arraybackslash}p{5.8cm}|}{\textbf{Command}} &
\multicolumn{1}{>{\centering\arraybackslash}p{4.2cm}|}{\textbf{Trigger Condition}} &
\multicolumn{1}{>{\centering\arraybackslash}p{4.0cm}|}{\textbf{Purpose}} \\ \hline

Linux &
\texttt{auditctl -w /var/lib/db/data.db -p wa -k db\_watch} &
Write or attribute change to database file &
Tag events as \texttt{db\_watch} for monitoring \\ \hline

Linux &
\texttt{auditctl -a always,exit -F arch=b64 -S openat,openat2 -F dir=/var/lib/db -F perm=w -F uid!=1001 -k oob\_db\_write} &
Write attempt by non-authorized UID &
Detect out-of-band writes from unauthorized processes \\ \hline

Linux &
\texttt{auditctl -e 2} &
Audit configuration changes &
Lock audit rules until reboot to prevent tampering \\ \hline

Windows &
\texttt{AuditPol /set /subcategory:"File System" /success:enable} &
Successful file write operation &
Enable auditing for writes; trigger Volume Shadow Copy for unauthorized writers \\ \hline

\end{tabular}
\end{adjustbox}
\vspace{-0.3cm}
\caption{Example snapshot-trigger configurations for Linux and Windows}
\label{table:snapshot_trigger_commands}
\end{table}

For example, Table~\ref{table:snapshot_trigger_commands} shows sample SA-owned configurations for Linux and Windows that detect both backdoored-application writes and out-of-band modifications. On Linux, immutable audit rules monitor the database file for any write or attribute-change events and tag them for monitoring. Additional rules flag write attempts from any process whose user identifier differs from that of the authorized service account, ensuring that only approved binaries can modify the database. The audit configuration is then locked until reboot to prevent tampering. On Windows, file-system auditing is enabled for successful write operations, producing Security Event~4663 entries. These events can be linked to a Volume Shadow Copy trigger when the writer’s security identifier does not match the authorized service account, ensuring that suspicious writes automatically generate a forensic snapshot.

These configurations are applied at the host layer under SA control, making them resilient to bypass attempts by a malicious DevOps Engineer without elevated privileges. Root or kernel-level attacks, such as forging inode change times, unloading kernel hooks, or writing directly to raw storage devices are outside the scope of this work and are mitigated through secure boot, kernel lockdown, and two-person control for administrative access.

\section{RADAR Framework \& Data Sources}
\label{sec:framework}

The RADAR framework provides an end-to-end process for detecting unattributed (unlogged) operations by correlating low-level storage artifacts with high-level audit logs. Its primary function is to identify discrepancies between the physical evidence carved from storage and the logical history recorded in an audit log, which may serve as strong indicators of unauthorized activity concealed by the adversary described in our threat model.

\subsection{Overall Architecture}
\label{sec:architecture}

\begin{figure*}[!t]
    \centering
    \includegraphics[width=1\textwidth]{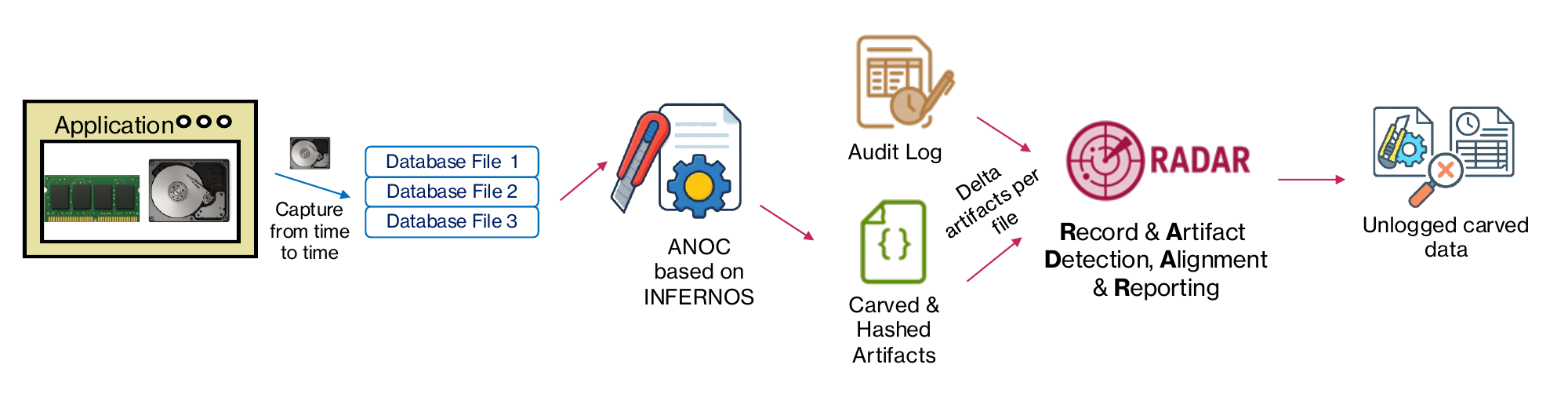}
    \caption{Architecture of RADAR}

    \label{fig:radar-workflow}
\end{figure*}

The RADAR workflow, illustrated in Figure~\ref{fig:radar-workflow}, follows a structured, multi-stage process. The process begins with the acquisition of a forensic snapshot of the database, which consists of the relevant database file or set of files captured automatically based on the host-level policy defined in Section~\ref{sec:threat-model}. These snapshots are then processed by the framework's carving engine to extract raw physical records. This crucial step is performed by our Automated NoSQL Carver (ANOC), a generalized tool designed to reconstruct database contents directly from raw disk or memory images, bypassing the database's API. As further detailed in Section~\ref{sec:acquisition}, ANOC is guided by the INFERNOS algorithm to automatically infer the storage layout from the raw bytes. This process yields a structured set of carved artifacts, which includes metadata essential for the reconciliation stage. For page-based databases like Berkeley DB, this includes page-level hashes used for comparative analysis, while for log-structured databases like ZODB, it includes historical versions of each record.

In parallel, the application-layer audit log is collected to serve as the official record of declared operations. Finally, the RADAR analysis engine ingests both the carved artifacts from ANOC and the audit log. As the core of the framework, RADAR performs a cross-validation by attempting to align each physical artifact with a corresponding logical operation in the log. Any carved record that cannot be explained by the log is flagged as an unattributed operation, or a ``delta artifact''. The specific data sources are detailed in Section \ref{sec:data-sources}, and the reconciliation logic is presented in Section \ref{sec:methodology}.

\subsection{Forensic Data Sources}
\label{sec:data-sources}
RADAR’s threat detection methodology reconciles two independent evidence streams: storage-level artifacts carved from forensic snapshots, which represent the \textbf{physical ground truth}, and an application-layer audit log, which records the \textbf{declared history of intended operations}. This section defines the log stream we accept and how we normalize it for analysis; the storage stream and carving process are covered in Section \ref{sec:acquisition}.

Specifically, Section \ref{sec:jsonl} explains the rationale for using an application-produced JSON Lines (JSONL) audit log and details the schema and canonicalization rules required to make each log entry comparable to a carved record. Finally, Section \ref{sec:excluded-wal} justifies the deliberate exclusion of engine-specific WALs and journals, as their heterogeneous, optional, and easily suppressed nature makes them unsuitable as a primary evidence source.

This approach allows the framework to remain both generalized and resilient. The carved storage provides log-independent ground truth, while the standardized application log supplies the declared history. RADAR’s analysis engine cross-checks the two, reporting any carved record that lacks a matching log explanation as an unattributed operation.

\subsubsection{Application Audit Log (JSONL)}
\label{sec:jsonl}

While relational databases often provide robust, built-in auditing, the NoSQL landscape presents a significant gap in security readiness. Many NoSQL databases lack a native audit logging capability entirely \cite{rajatheva2019comparative, tawalbeh2013security}. In other prominent systems, this critical security feature is positioned only as a premium, enterprise-level add-on, leaving users of free or community versions without a built-in solution. As a result, the absence of an accessible, standardized audit mechanism forces the responsibility of generating a secure and verifiable log back to the application layer.

To fill this void, the industry has widely adopted JSON Lines (JSONL) as a de facto standard for creating a portable and transparent audit trail \cite{Kleppmann2017}. Consequently, our work employs the JSONL format, aligning our methodology with established practices in security monitoring. Under our threat model, this application audit is treated as the authoritative record of intended operations, and engine-managed logs such as WALs or journals are excluded. JSONL is well-suited for this purpose. Its stream-oriented, append-only nature is highly efficient for continuous data flows. For data integrity, its line-by-line structure, where each line is a self-contained object, makes log files more resilient to corruption than monolithic JSON files \cite{Kleppmann2017, JSONLinesOrg}. For compliance, this format is ideal for producing the context-rich event records demanded by regulations like SOX and HIPAA \cite{HHSGovHIPAA, Kent2007NIST}.

\paragraph{Schema}
Each log line $e$ is a single JSON object containing six fields:
\[
e=\{\texttt{ts},\,\texttt{op},\,\texttt{key},\,\texttt{user},\,\texttt{old\_value},\,\texttt{new\_value}\},
\]
where \texttt{op} specifies the \textbf{operation} type, which is one of $\{\texttt{insert},\texttt{delete},\texttt{update},\\\texttt{update\_fields},\texttt{delete\_fields}\}$. The \textbf{timestamp} \texttt{ts} advances monotonically for each writer: every new entry must have a time equal to or later than the previous one, with ties allowed when multiple operations occur within the same second. For example, two inserts committed in a batch may appear with identical timestamps or with consecutive second-level timestamps (e.g., 12:00:01, 12:00:02). Timestamps serve as ordering aids only, since reconciliation relies on key-and-value agreement rather than time alone. We support both UNIX epoch (integer seconds since 1970) and RFC3339 (ISO-like string) encodings. The \texttt{user} field contains the identifier of the user who performed the action and supports attribution, though it is not required for authorization. The \texttt{key} field holds the record's unique identifier.

In key-value stores, the \texttt{old\_value} and \texttt{new\_value} fields contain the state of the record before and after the operation as scalar strings (e.g., \texttt{"CANADA\_00001"}). In document stores, these values may appear as native JSON objects or as serialized JSON strings with escaped quotes (e.g., \verb|{"_id":"part#000002000",...}|); both encodings are supported.

\paragraph{Canonicalization}
Values are compared using a deterministic canonicalization function $\mathrm{canon}(\cdot)$ to avoid spurious mismatches from JSON key order, whitespace, or storage metadata. When a value arrives as a JSON string, the canonicalization step parses it and converts it to canonical JSON with sorted keys and minimal whitespace. When a value arrives as a JSON object, the same step converts it directly to canonical JSON. Scalar strings use byte-for-byte equality. Key normalization (e.g., case folding) may be applied before matching. Engine metadata fields such as \texttt{status}, \texttt{page\_id}, \texttt{page\_MD5}, and \texttt{page\_offset} are excluded from equality tests to prevent false positives caused by layout or page-hash changes.\footnote{This mirrors forensic practice that avoids flagging purely physical rearrangements as logical edits.} For document stores, the comparison uses whole canonical documents. Field-level diffs are not used unless the audit itself is field-level (see Section~\ref{sec:reconcile}). In that case, matches are labeled “authorized (field-level)” in the results.

\subsubsection{Excluded Engine-Specific Logs (WAL/Journals)}
\label{sec:excluded-wal}
Although the WAL is a forensically valuable artifact, it is deliberately excluded as a primary evidence source in the \radar{} framework for several reasons. First, WAL formats across NoSQL systems are often proprietary, binary, and undocumented, requiring specialized, database-specific tools for reliable parsing. Developing and maintaining such parsers for a general-purpose framework is outside the scope of this research. Second, many NoSQL databases, such as LMDB and MDBX, do not implement a WAL at all, while others, including ZODB and Berkeley DB, permit journaling or transactional logging to be disabled or bypassed under certain configurations, such as for performance optimization or bulk data operations. Even in engines that implement transactional logging for durability, such as RavenDB with its Write-Ahead Journal or etcd, which uses a Raft log to replicate and persist operations across cluster nodes, a privileged administrator can force log rotation and delete older log files, erasing evidence of unauthorized actions. Additionally, LiteDB and RealmDB do not provide a traditional, persistent write-ahead log, while NitriteDB, although supporting some logging, relies on application-level logs that are non-persistent and lack tamper resistance. Given these complexities and the ease with which WAL artifacts can be absent, altered, or suppressed, the \radar{} framework instead emphasizes analysis of persistent storage artifacts, which provide a more universally applicable and resilient basis for malicious activity detection.

\subsection{Acquisition \& Carving (ANOC)}
\label{sec:acquisition}

RADAR's analysis begins with carved data generated by the \texttt{Automated NoSQL Carver (ANOC)}. As detailed in our previous work \cite{nissan2025anoc}, ANOC is a novel tool designed to reconstruct database contents from raw bytes without relying on database APIs or logs. It operates in two phases. The first, a one-time \emph{Parameter Collector}, utilizes the \texttt{INFERNOS} (Inference for NoSQL Storage) algorithm. INFERNOS automates the reverse-engineering of a database’s storage architecture by applying a series of statistical analyses and heuristics to the raw binary bytes. It infers critical parameters such as page size, record directory, and record-encoding formats without prior knowledge of the specific database type. The second phase is a read-only \emph{Carver} that applies the structural profile generated by INFERNOS to extract records from a target image.

RADAR leverages this carved output to create a uniform logical view of the database state for reconciliation against the audit log. This view provides three essential components for each carved item: a \texttt{correlation key} compatible with the \texttt{key} field in the audit log, a \texttt{canonicalized value} for accurate content comparison, and a \texttt{status indicator} (e.g., `active' or `deleted') to distinguish current records from historical remnants. While RADAR's attribution logic relies on these three components, it also uses other carved metadata, such as page-level hashes for in-place modification databases, to perform preliminary filtering and efficiently identify modified pages.

ANOC's reliance on a learned profile, rather than hardcoded layouts, makes it highly extensible and has been successfully applied to ten representative NoSQL engines, including key-value stores (Berkeley~DB, LMDB, MDBX, ZODB, Durus, etcd) and document stores (RavenDB, LiteDB, Realm, NitriteDB). The carver emits this data through a standardized interface, a design choice that keeps RADAR decoupled from engine internals. This carved output is the sole input RADAR requires from storage. For engines that preserve historical versions, RADAR applies a single-snapshot analysis (Section \ref{sec:method1}), while for those that perform in-place updates, it uses a comparative analysis of a \emph{before} and \emph{after} snapshot pair (Section \ref{sec:method2}). In all cases, the final reconciliation proceeds against the application audit log using the decision rules detailed in Section \ref{sec:reconcile}.

\section{Threat Detection Methodology}
\label{sec:methodology}
\subsection{Forensic Artifacts in NoSQL Storage}
\label{sec:artifacts}

The forensic analysis of NoSQL systems is built upon a fundamental principle: every logical database operation leaves a corresponding physical footprint in the persistent storage layer. Because these low-level alterations cannot be easily bypassed, they serve as a reliable ground truth for verification. The principle of \emph{storage remanence} ensures that residual artifacts persist until they are physically overwritten, and these remnants can be recovered through carving. The persistence and recoverability of these artifacts, however, depend critically on the engine's underlying storage architecture. These architectures fall into two categories: systems using append-only and copy-on-write (CoW), and systems using in-place modification.

\paragraph{Append-Only and Copy-on-Write (CoW) Systems} These engines handle modifications by writing a new version of a record or data block, leaving the original data physically intact. This approach preserves a full historical record of prior states within the database file itself. Deleted records are not erased but are simply superseded by newer entries, making them directly recoverable from a single forensic snapshot.

\paragraph{In-Place Modification Systems}
In contrast, these engines apply updates by overwriting data within an existing record slot or page, so prior values are not durably retained as addressable artifacts in the database file. Deletions typically invalidate a slot or return space to the allocator (marking it \emph{free}). Although the old bytes may temporarily remain in unallocated space, they are unreferenced and may be overwritten at any time. Therefore, \radar{} does not rely on these uncertain artifacts.

Consequently, to address this distinction in artifact preservation, \radar{} employs two distinct methodologies, each specifically designed for the characteristics of the underlying storage architecture: a single-snapshot analysis for systems that preserve history (\texttt{Method 1}), and a comparative analysis for those that do not (\texttt{Method 2}). Sections \ref{sec:method1} and \ref{sec:method2} detail these approaches and the reconciliation rules used to detect unattributed operations by cross-referencing carved artifacts with audit logs

\subsection{Method 1: Single-Snapshot Analysis for Append-Only \& CoW Databases}
\label{sec:method1}

For storage architectures that preserve historical versions, such as append-only or Copy-on-Write (CoW) systems, \radar{} can detect unattributed operations by analyzing a single forensic snapshot. This snapshot contains both active and historical records, allowing for a comprehensive reconstruction of the database's state transitions. The goal of this method is to verify that every transition is justified by an entry in the application audit log. The analysis proceeds in three stages, formalized in Algorithms \ref{alg:detect-deletes}, \ref{alg:detect-inserts}, and \ref{alg:detect-updates}.

First, the framework identifies all carved records with a ``Deleted'' status and flags any that cannot be matched to a corresponding `delete' operation or the `old value' of an `update' operation in the log as an \texttt{Unattributed Delete}. Second, it examines all ``Active'' records and flags any that do not correspond to a logged `insert' or the `new value' of an `update' as an \texttt{Unattributed Insert}. Finally, because a single unlogged update generates both an unattributed deleted record (the old value) and an unattributed active record (the new value) for the same key, a third algorithm correlates these findings, consolidating such pairs into a single \texttt{Unattributed Update}.

\subsubsection{Detecting Unattributed Deletes}

The first stage of the analysis identifies records that were marked as deleted in the database but whose deletion is not documented in the audit log. The detection process cross-references these carved remnants against the log to find discrepancies, as formalized in Algorithm~\ref{alg:detect-deletes}. For a key-value store, the value is typically a scalar; for a document store, it is the entire document object. Since an update operation in these systems effectively deletes the old version of a document, the algorithm validates the entire state of the data at the time of deletion against both `delete' and `update' operations in the log.

\begin{algorithm}[!ht]
\SetAlgoLined
\SetKwInOut{Input}{Input}
\SetKwInOut{Output}{Output}
\Input{$C$: carved record set}
\Input{$L$: audit log}
\KwOut{$\Delta_{\mathrm{del}}$: unattributed deletes}

$\Delta_{\mathrm{del}} \leftarrow \emptyset$\; \label{alg:del:init_delta}
$S \leftarrow \emptyset$\; \label{alg:del:init_s}
\ForEach{record $r \in C$}{ \label{alg:del:loop}
  \If{$r.\texttt{status} = \texttt{Delete} \land (r.\texttt{key}, r.\texttt{value}) \notin S \land \neg\exists\,l \in L:((l.\texttt{op} = \texttt{delete} \land l.\texttt{key} = r.\texttt{key} \land l.\texttt{value} = r.\texttt{value}) \lor (l.\texttt{op} = \texttt{update} \land l.\texttt{key} = r.\texttt{key} \land l.\texttt{old\_value} = r.\texttt{value}))$}{ \label{alg:del:if}
    $\Delta_{\mathrm{del}} \leftarrow \Delta_{\mathrm{del}} \cup \{r\}$\; \label{alg:del:add_delta}
    $S \leftarrow S \cup \{(r.\texttt{key}, r.\texttt{value})\}$\; \label{alg:del:add_s}
  }
}
\Return{$\Delta_{\mathrm{del}}$}\; \label{alg:del:return}
\caption{Detect Unattributed Deletes}
\label{alg:detect-deletes}
\end{algorithm}
\FloatBarrier

The algorithm receives as input the carved record set $C$ and the audit log $L$, and produces the set $\Delta_{\mathrm{del}}$ containing all deleted records that lack attribution in the log. The sets $\Delta_{\mathrm{del}}$ and $S$ are initialized as empty (lines~\ref{alg:del:init_delta}--\ref{alg:del:init_s}), where $S$ is used to track processed key-value pairs and prevent duplicate reporting. The main loop iterates through each carved record $r$ in $C$ (line~\ref{alg:del:loop}). For each record, the conditional in line~\ref{alg:del:if} verifies that the record's status is \texttt{Delete}, its $(\texttt{key}, \texttt{value})$ pair has not already been processed (i.e., is not in $S$), and there does not exist any log entry $l$ in $L$ that can account for the deletion. Specifically, attribution is only established if there is either an explicit \texttt{delete} operation in the log with the same key and value, or an \texttt{update} operation with the same key and a matching \texttt{old\_value}. If neither condition holds, the record $r$ is added to $\Delta_{\mathrm{del}}$ (line~\ref{alg:del:add_delta}), and its $(\texttt{key}, \texttt{value})$ pair is added to $S$ (line~\ref{alg:del:add_s}) to ensure uniqueness in subsequent iterations. Once all records in $C$ have been evaluated, the algorithm returns the set $\Delta_{\mathrm{del}}$ (line~\ref{alg:del:return}), representing the unique deleted records for which no corresponding action exists in the audit log.

For key-value stores, the \texttt{value} field typically holds a scalar. For document stores, the value is the complete document (e.g., a JSON object), so the algorithm ensures that the entirety of the deleted state is compared for attribution. This approach supports precise forensic reconstruction by ensuring that only unexplained, non-redundant deleted records are reported as unattributed.

\begin{figure*}[!ht]
    \centering
	\includegraphics[width=0.8\textwidth]{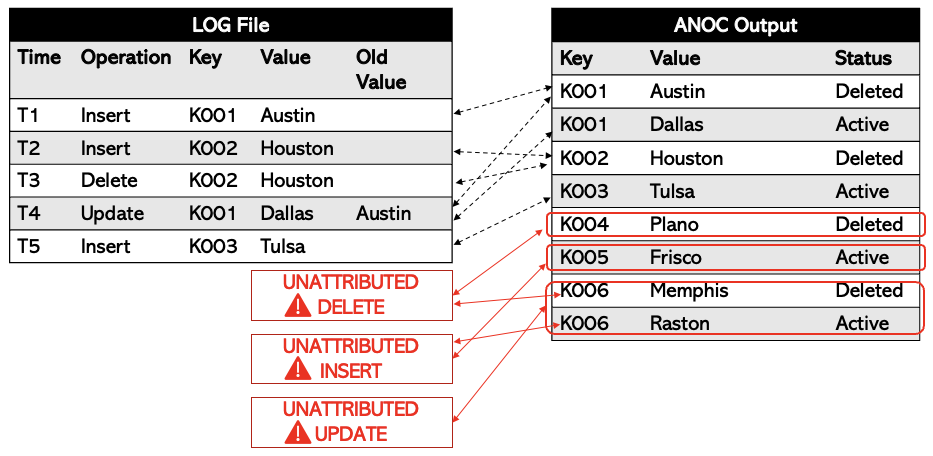}
    \caption{Unattributed Operations (Key-Value Store)}	
	\label{fig:unattributed-kv}
\end{figure*}

\begin{figure*}[!ht]
    \centering
	\includegraphics[width=0.8\textwidth]{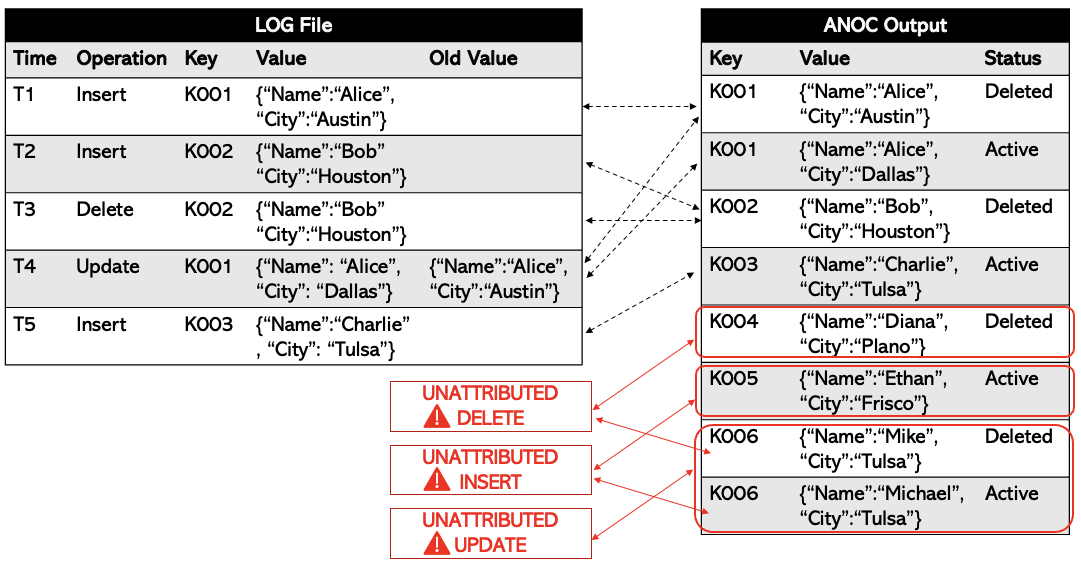}
    \caption{Unattributed Operations (Document Store)}	
	\label{fig:unattributed-doc}
\end{figure*}

\paragraph{Example (Key-Value Store)}

As shown in Figure~\ref{fig:unattributed-kv}, the audit log and ANOC output are cross-referenced to identify unattributed deletes. The deleted records for (\texttt{K001}, \texttt{Austin}) and (\texttt{K002}, \texttt{Houston}) are explained by matching log entries. However, the deleted records (\texttt{K004}, \texttt{Plano}) and (\texttt{K006}, \texttt{Memphis}) are not supported by any log entry, as there are no corresponding delete or update (\texttt{old\_value}) operations for them. According to Algorithm~\ref{alg:detect-deletes}, they are therefore reported as unattributed deletes.

\paragraph{Example (Document Store)}

In the document store scenario (see Figure~\ref{fig:unattributed-doc}), each value is a complete document. The deleted records for \texttt{K001} and \texttt{K002} are accounted for by the log. However, the deleted documents (\texttt{K004}, \texttt{\{"Name":"Diana", "City":"Plano"\}}) and (\texttt{K006}, \texttt{\{"Name":"Mike", "City":"Tulsa"\}}) are not explained by any log entry. As a result, the algorithm flags these documents as unattributed deletes.

\subsubsection{Detecting Unattributed Inserts}

The second stage of the analysis focuses on ``active'' records carved by ANOC. These records may represent legitimate insertions, the after-state of an update, or records created by an unlogged malicious process. To distinguish between these possibilities, it is essential to cross-reference each carved active record with both `insert` and `update` operations in the audit log. An active record is flagged as an unattributed insert only when it cannot be explained by either type of logged operation, as formalized in Algorithm~\ref{alg:detect-inserts}.

\begin{algorithm}[!ht]
\SetAlgoLined
\SetKwInOut{Input}{Input}
\SetKwInOut{Output}{Output}
\Input{$C$: carved record set}
\Input{$L$: audit log}
\KwOut{$\Delta_{\mathrm{ins}}$: unattributed inserts}

$\Delta_{\mathrm{ins}} \leftarrow \emptyset$\; \label{alg:ins:init_delta}
$S \leftarrow \emptyset$\; \label{alg:ins:init_s}
\ForEach{record $r \in C$}{ \label{alg:ins:loop}
  \If{$r.\texttt{status} = \texttt{Active} \land (r.\texttt{key}, r.\texttt{value}) \notin S \land \neg\exists\,l \in L:((l.\texttt{op} = \texttt{insert} \land l.\texttt{key} = r.\texttt{key} \land l.\texttt{value} = r.\texttt{value}) \lor (l.\texttt{op} = \texttt{update} \land l.\texttt{key} = r.\texttt{key} \land l.\texttt{new\_value} = r.\texttt{value}))$}{ \label{alg:ins:if}
    $\Delta_{\mathrm{ins}} \leftarrow \Delta_{\mathrm{ins}} \cup \{r\}$\; \label{alg:ins:add_delta}
    $S \leftarrow S \cup \{(r.\texttt{key}, r.\texttt{value})\}$\; \label{alg:ins:add_s}
  }
}
\Return{$\Delta_{\mathrm{ins}}$}\; \label{alg:ins:return}
\caption{Detect Unattributed Inserts}
\label{alg:detect-inserts}
\end{algorithm}
\FloatBarrier

The algorithm receives as input the carved record set $C$ and the audit log $L$, and produces the set $\Delta_{\mathrm{ins}}$ containing all active records that lack attribution in the log. The sets $\Delta_{\mathrm{ins}}$ and $S$ are initialized as empty (lines~\ref{alg:ins:init_delta}--\ref{alg:ins:init_s}), where $S$ is used to track processed key-value pairs and prevent duplicate reporting. The main loop iterates through each carved record $r$ in $C$ (line~\ref{alg:ins:loop}). For each record, the conditional in line~\ref{alg:ins:if} verifies that the record's status is \texttt{Active}, its $(\texttt{key}, \texttt{value})$ pair has not already been processed (i.e., is not in $S$), and there does not exist any log entry $l$ in $L$ that can account for the insertion. Specifically, attribution is only established if there is either an explicit \texttt{insert} operation in the log with the same key and value, or an \texttt{update} operation with the same key and a matching \texttt{new\_value}. If neither condition holds, the record $r$ is added to $\Delta_{\mathrm{ins}}$ (line~\ref{alg:ins:add_delta}), and its $(\texttt{key}, \texttt{value})$ pair is added to $S$ (line~\ref{alg:ins:add_s}) to ensure uniqueness in subsequent iterations. Once all records in $C$ have been evaluated, the algorithm returns the set $\Delta_{\mathrm{ins}}$ (line~\ref{alg:ins:return}), representing the unique active records for which no corresponding action exists in the audit log.

For key-value stores, the \texttt{value} field typically holds a scalar. For document stores, the value is the complete document (e.g., a JSON object), so the algorithm ensures that the entirety of the active state is compared for attribution. This approach supports precise forensic reconstruction by ensuring that only unexplained, non-redundant active records are reported as unattributed.

\paragraph{Example (Key-Value Store)}
As shown in Figure~\ref{fig:unattributed-kv}, the ANOC output is cross-referenced with the audit log to identify unexplained active records. Each (\texttt{key}, value) pair with \texttt{Active} status is checked for a matching \texttt{insert} or \texttt{update} (\texttt{new\_value}) operation in the log. While records such as (\texttt{K001}, \texttt{Dallas}) and (\texttt{K003}, \texttt{Tulsa}) are explained by the audit log, the active records (\texttt{K005}, \texttt{Frisco}) and (\texttt{K006}, \texttt{Raston}) have no corresponding log entries. As specified by Algorithm~\ref{alg:detect-inserts}, these records are therefore flagged as \emph{unattributed inserts}.

\paragraph{Example (Document Store)}
As illustrated in Figure~\ref{fig:unattributed-doc}, the ANOC output reconstructs a set of active documents from storage. Each active (\texttt{key}, document) pair is compared to the audit log to determine if its presence is supported by a corresponding \texttt{insert} or \texttt{update} (\texttt{new\_value}) operation. While most active documents are explained by matching log entries, the records (\texttt{K005}, \{\texttt{"Name":"Ethan", "City":"Frisco"}\}) and (\texttt{K006}, \{\texttt{"Name":"Michael", "City":"Tulsa"}\}) appear as active records in storage but lack any corresponding operation in the log. According to Algorithm~\ref{alg:detect-inserts}, these records are reported as \emph{unattributed inserts}.

\subsubsection{Detecting Unattributed Updates}

In modern NoSQL systems,  especially those using append-only or copy-on-write architectures an \texttt{UPDATE} operation typically leaves two forensic artifacts for the same key: a deleted “before” version (old value) and an active “after” version (new value). After the initial analysis has identified all unattributed deletes and inserts, the framework's final step is to correlate these findings. Consequently, an unattributed update is defined as a pair, consisting of a record from $\Delta_{\mathrm{del}}$ and a record from $\Delta_{\mathrm{ins}}$, that share the same key but have different values. This approach isolates update pairs by correlating the initial result sets without re-scanning the much larger carved set $C$. Algorithm~\ref{alg:detect-updates} outlines the procedure for matching these update pairs.

\begin{algorithm}[!ht]
\SetAlgoLined
\SetKwInOut{Input}{Input}
\SetKwInOut{Output}{Output}
\Input{$\Delta_{\mathrm{del}}$: unattributed deletes}
\Input{$\Delta_{\mathrm{ins}}$: unattributed inserts}
\KwOut{$R_{\mathrm{del}}, R_{\mathrm{ins}}, R_{\mathrm{upd}}$: final disjoint sets}

$R_{\mathrm{del}} \leftarrow \Delta_{\mathrm{del}}$\; \label{alg:upd:init_rdel}
$R_{\mathrm{ins}} \leftarrow \Delta_{\mathrm{ins}}$\; \label{alg:upd:init_rins}
$R_{\mathrm{upd}} \leftarrow \emptyset$\; \label{alg:upd:init_rupd}

\tcp{Build a map of active records for fast lookup}
ins\_map $\leftarrow$ \texttt{new Map<Key, List<Record>>}\; \label{alg:upd:insmap}
\ForEach{record $r \in R_{\mathrm{ins}}$}{
    add $r$ to $ins\_map[r.\texttt{key}]$\;
}

\tcp{Find and consolidate update pairs}
\ForEach{record $r_{\mathrm{del}}$ in $\Delta_{\mathrm{del}}$}{ \label{alg:upd:outerloop}
    \If{$ins\_map$ contains $r_{\mathrm{del}}.\texttt{key}$}{ \label{alg:upd:ifmap}
        \ForEach{record $r_{\mathrm{act}}$ in $ins\_map[r_{\mathrm{del}}.\texttt{key}]$}{ \label{alg:upd:innerloop}
            \If{$r_{\mathrm{del}}.\texttt{value} \neq r_{\mathrm{act}}.\texttt{value}$}{ \label{alg:upd:valcheck}
                $R_{\mathrm{upd}} \leftarrow R_{\mathrm{upd}} \cup \{(r_{\mathrm{del}}, r_{\mathrm{act}})\}$\; \label{alg:upd:addpair}
                $R_{\mathrm{del}} \leftarrow R_{\mathrm{del}} \setminus \{r_{\mathrm{del}}\}$\; \label{alg:upd:removedel}
                $R_{\mathrm{ins}} \leftarrow R_{\mathrm{ins}} \setminus \{r_{\mathrm{act}}\}$\; \label{alg:upd:removeins}
                break\; \label{alg:upd:break}
            }
        }
    }
}
\Return{$R_{\mathrm{del}}, R_{\mathrm{ins}}, R_{\mathrm{upd}}$}\; \label{alg:upd:return}
\caption{Detect Unattributed Updates}
\label{alg:detect-updates}
\end{algorithm}
\FloatBarrier

It processes the sets of unattributed deletes and inserts to produce three mutually exclusive sets: $R_{\mathrm{del}}$ (final unexplained deletes), $R_{\mathrm{ins}}$ (final unexplained inserts), and $R_{\mathrm{upd}}$ (update pairs). The sets $R_{\mathrm{del}}$ and $R_{\mathrm{ins}}$ are initialized as copies of $\Delta_{\mathrm{del}}$ and $\Delta_{\mathrm{ins}}$, respectively, and $R_{\mathrm{upd}}$ is initialized as empty (lines~\ref{alg:upd:init_rdel}--\ref{alg:upd:init_rupd}). For efficient lookup, an auxiliary map $ins\_map$ is created and populated to group the active insert records by their key. This enables rapid lookup of potential update candidates for each deleted record.

The main loop iterates over each record $r_{\mathrm{del}}$ in $\Delta_{\mathrm{del}}$ (line~\ref{alg:upd:outerloop}). For each candidate delete, the algorithm checks whether $ins\_map$ contains any active record with the same key (line~\ref{alg:upd:ifmap}). If so, it examines each corresponding active record $r_{\mathrm{act}}$ (line~\ref{alg:upd:innerloop}), and compares their values (line~\ref{alg:upd:valcheck}). If the values are different, the two records constitute a candidate update pair: this pair $(r_{\mathrm{del}}, r_{\mathrm{act}})$ is added to $R_{\mathrm{upd}}$ (line~\ref{alg:upd:addpair}), and both $r_{\mathrm{del}}$ and $r_{\mathrm{act}}$ are removed from $R_{\mathrm{del}}$ and $R_{\mathrm{ins}}$, respectively (lines~\ref{alg:upd:removedel}–\ref{alg:upd:removeins}), to maintain the disjoint property. The inner loop is terminated immediately (line~\ref{alg:upd:break}) to enforce a one-to-one mapping per key.

After all possible update pairs have been identified, the algorithm returns the refined, mutually exclusive sets of unexplained deletes, inserts, and updates (line~\ref{alg:upd:return}). This approach ensures that each unexplained update is attributed to exactly one deleted and one active record with the same key but different values, enabling clear forensic separation of unexplained modifications.

\paragraph{Example (Key-Value Store)}
The same figure illustrates the process for identifying \emph{unattributed updates}. In Figure~\ref{fig:unattributed-kv}, the pair (\texttt{K006}, \texttt{Memphis}) [deleted] and (\texttt{K006}, \texttt{Raston}) [active] share the same key but have different values. The initial detection flags both as individual unattributed findings. As outlined in Algorithm~\ref{alg:detect-updates}, the framework then consolidates these two findings into a single unattributed update pair, revealing a state transition that cannot be explained by the audit log.

\paragraph{Example (Document Store)}
The same figure demonstrates the process for consolidating \emph{unattributed updates}. In Figure~\ref{fig:unattributed-doc}, the pair (\texttt{K006}, \{\texttt{"Name":"Mike", "City":"Tulsa"}\}) [deleted] and (\texttt{K006}, \{\texttt{"Name":"Michael", "City":"Tulsa"}\}) [active] share the same key but have different values. As described in Algorithm~\ref{alg:detect-updates}, this combination is flagged as an \emph{unattributed update}, reflecting an unlogged or unauthorized change that is detectable only via forensic carving.

\subsection{Method 2: Comparative Analysis for In-Place Modification Databases}
\label{sec:method2}

For storage engines that perform in-place modifications, historical versions of records are overwritten and destroyed, making the single-snapshot analysis of Method 1 ineffective. To detect unattributed operations in such cases, \radar{} employs a \texttt{comparative analysis} methodology. This approach requires two forensic snapshots carved by ANOC: a trusted baseline (``before'') state and a suspicious (``after'') state. By comparing these snapshots at the page level, \radar{} can identify physical changes and then correlate them with the audit log to find discrepancies. The comparative analysis process, formally defined in Algorithm \ref{alg:compare-snapshots}, begins by comparing the MD5 hash of each page between the two snapshots. A hash mismatch triggers a detailed, record-level analysis of that page to determine if the detected changes, e.g., updates, inserts, or deletes have corresponding legitimate log entries. If no such entry exists, the operation is flagged as unattributed. This method allows \radar{} to detect unlogged activities even in databases that destroy historical evidence.

\begin{algorithm}[!ht]
\SetAlgoLined
\SetKwInOut{Input}{Input}
\SetKwInOut{Output}{Output}
\SetKwFunction{ParseLog}{ParseAuditLog}
\SetKwFunction{GetPages}{GetPages}
\SetKwFunction{GetPage}{GetPageByIndex}
\SetKwFunction{GetHash}{GetPageHash}
\SetKwFunction{GetRecords}{GetRecordsOnPage}
\SetKwFunction{CheckLog}{IsOperationLogged}

\Input{$S_{\mathrm{before}}$: Carved snapshot of the ``before'' state}
\Input{$S_{\mathrm{after}}$: Carved snapshot of the ``after'' state}
\Input{$L$: Audit log}
\KwOut{$U_{\mathrm{updates}}, U_{\mathrm{inserts}}, U_{\mathrm{deletes}}$: Sets of unattributed operations}

$L_{\mathrm{map}} \leftarrow \ParseLog(L)$\; \label{alg:cs:parse_log}
$U_{\mathrm{updates}}, U_{\mathrm{inserts}}, U_{\mathrm{deletes}} \leftarrow \emptyset, \emptyset, \emptyset$\;

\ForEach{page $p_{\mathrm{before}}$ in \GetPages{$S_{\mathrm{before}}$}}{ \label{alg:cs:page_loop}
    $p_{\mathrm{after}} \leftarrow \GetPage(S_{\mathrm{after}}, p_{\mathrm{before}}.\texttt{index})$\;
    
    \tcp{Analyze page only if its content has changed}
    \If{\GetHash{$p_{\mathrm{before}}$} $\neq$ \GetHash{$p_{\mathrm{after}}$}}{ \label{alg:cs:hash_check}
        $R_{\mathrm{before}} \leftarrow \GetRecords(p_{\mathrm{before}})$\;
        $R_{\mathrm{after}} \leftarrow \GetRecords(p_{\mathrm{after}})$\;
        $K \leftarrow R_{\mathrm{before}}.\texttt{keys} \cup R_{\mathrm{after}}.\texttt{keys}$\; \label{alg:cs:union_keys}
        
        \ForEach{key $k \in K$}{ \label{alg:cs:key_loop}
            $r_{\mathrm{b}}, r_{\mathrm{a}} \leftarrow R_{\mathrm{before}}[k], R_{\mathrm{after}}[k]$\;
            \uIf{$r_{\mathrm{b}}$ exists and $r_{\mathrm{a}}$ exists and $r_{\mathrm{b}}.\texttt{value} \neq r_{\mathrm{a}}.\texttt{value}$}{ \label{alg:cs:update_check}
                \If{not \CheckLog{$L_{\mathrm{map}}$, $k$, `update'}}{
                    $U_{\mathrm{updates}} \leftarrow U_{\mathrm{updates}} \cup \{(k, r_{\mathrm{b}}, r_{\mathrm{a}})\}$\;
                }
            }
            \uElseIf{$r_{\mathrm{b}}$ exists and $r_{\mathrm{a}}$ does not exist}{ \label{alg:cs:delete_check}
                \If{not \CheckLog{$L_{\mathrm{map}}$, $k$, `delete'}}{
                    $U_{\mathrm{deletes}} \leftarrow U_{\mathrm{deletes}} \cup \{(k, r_{\mathrm{b}})\}$\;
                }
            }
            \uElseIf{$r_{\mathrm{a}}$ exists and $r_{\mathrm{b}}$ does not exist}{ \label{alg:cs:insert_check}
                 \If{not \CheckLog{$L_{\mathrm{map}}$, $k$, `insert'}}{
                    $U_{\mathrm{inserts}} \leftarrow U_{\mathrm{inserts}} \cup \{(k, r_{\mathrm{a}})\}$\;
                }
            }
        }
    }
}
\Return{$U_{\mathrm{updates}}, U_{\mathrm{inserts}}, U_{\mathrm{deletes}}$}\;
\caption{Analysis for In-Place Modification}
\label{alg:compare-snapshots}
\end{algorithm}

The algorithm's methodology is detailed as follows. It commences by parsing the entire audit log into a key-indexed map, $L_{\mathrm{map}}$, to facilitate efficient lookups (line~\ref{alg:cs:parse_log}). It also initializes empty sets to store any discovered unattributed updates, inserts, or deletes.

The core of the analysis involves iterating through each page, $p_{\mathrm{before}}$, in the baseline snapshot $S_{\mathrm{before}}$ (line~\ref{alg:cs:page_loop}). For each page, its corresponding page, $p_{\mathrm{after}}$, is retrieved from the ``after'' snapshot $S_{\mathrm{after}}$ by its index. A crucial optimization is performed in line~\ref{alg:cs:hash_check}: the algorithm compares the MD5 hash of $p_{\mathrm{before}}$ with that of $p_{\mathrm{after}}$. A granular, record-level analysis is initiated only if the hashes differ, indicating that the page's content has been modified. This page-level hashing dramatically reduces the scope of the investigation to only those pages with evident changes.

Upon detecting a hash mismatch, the algorithm proceeds to a detailed forensic examination. It extracts all records from both the ``before'' and ``after'' versions of the page, storing them in sets $R_{\mathrm{before}}$ and $R_{\mathrm{after}}$. To ensure no operation is missed, it creates a union of all record keys present in either set (line~\ref{alg:cs:union_keys}). It then iterates through each unique key $k$ (line~\ref{alg:cs:key_loop}) to assess its state transition. There are three possible changes for any given key:
\begin{itemize}
    \item \textbf{Update:} If a record with key $k$ exists in both snapshots but its value has changed (line~\ref{alg:cs:update_check}), the algorithm queries the log map to verify if a legitimate `update' operation was recorded. If not, the change is flagged as an unattributed update, capturing its key and both its before and after values.
    \item \textbf{Delete:} If a record with key $k$ exists in the baseline snapshot but is absent from the ``after'' snapshot (line~\ref{alg:cs:delete_check}), it signifies a deletion. The algorithm checks the log for a corresponding `delete' entry. If none is found, the operation is added to the set of unattributed deletes.
    \item \textbf{Insert:} If a record with key $k$ is present in the ``after'' snapshot but not in the baseline (line~\ref{alg:cs:insert_check}), it is identified as an insertion. The log is checked for an `insert' operation for that key. An absence of such a log entry results in the operation being classified as an unattributed insert.
\end{itemize}

Finally, after analyzing all modified pages, the algorithm returns the three sets containing all identified unattributed updates, inserts, and deletes, providing a comprehensive account of unlogged activities.

\subsection{Audit–Forensic Reconciliation \& Attribution (Decision Rules)}
\label{sec:reconcile}

In its final stage, reconciliation, \radar{} systematically cross-references the low-level evidence carved by ANOC against the high-level application audit log. This process applies a set of formal decision rules that attribute each physical state change to a logged operation and flags any artifact it cannot reconcile as an unattributed operation for forensic review. Two key principles form the foundation of \radar{}'s matching logic. First, to prevent false positives from insignificant variations in whitespace or key ordering in structured data like JSON, the framework normalizes all values into a canonical string format before comparison. Second, \radar{}'s analysis logic generates a content-only ``fingerprint'' of each record, explicitly excluding storage-layer metadata fields such as page identifiers or on-disk offsets. This design prevents the framework from misidentifying routine database operations, like compaction, as logical data modifications.

With these matching semantics, the framework applies its attribution logic. It flags an \texttt{Unattributed Delete} if a record is absent in a later state but its canonical value does not match a logged `delete' operation or the `old\_value' of a corresponding `update' operation. Conversely, it flags an \texttt{Unattributed Insert} if a new record appears but its value does not correspond to a logged `insert' or the `new\_value' of an `update'. The framework identifies an \texttt{Unattributed Update} either by correlating a delete and insert for the same key in single-snapshot analysis or by detecting a change in a record's content fingerprint between two snapshots where no `update' operation is logged. The framework's reliance on state-based reconciliation using keys and canonical values instead of timestamps makes it inherently resilient to operational variances such as clock skew and batch operations. This design is fundamental to its ability to function within the primary threat model of incomplete or suppressed logs, as it treats the carved storage as the ultimate source of ground truth.

\if false

\section{Trust in NoSQL Audit Logs}

NoSQL databases exhibit substantial variability in their audit logging mechanisms. Unlike relational databases that often employ tamper-resistant write-ahead logs (WAL) and centralized auditing, many NoSQL engines delegate logging to application or infrastructure-level components. These logs are typically recorded in plaintext or JSON formats and stored either as files on disk or forwarded to external services. Their integrity is governed by deployment-time configurations---precisely the scope under the control of a malicious DevOps Engineer (DO). This creates several vulnerabilities:

\subsubsection*{Suspending and Redirecting Logs} A DO can modify deployment configurations to disable or redirect logging. For example, in a Kubernetes environment, a DO can push a minor update to a deployment manifest (\texttt{deployment.yaml}) that changes a single line to redirect a log volume to \texttt{/dev/null}. This change is applied via a standard \texttt{kubectl apply -f} command, which orchestrators see as a routine "rolling update." Similarly, a Docker Compose file can be altered to change a container's \texttt{logging driver} to \texttt{none}, or a \texttt{systemd} service file can be modified to pipe \texttt{stdout} to \texttt{/dev/null}. Malicious actions can be performed during this window, after which normal logging is restored, leaving a gap that is difficult to detect.

\subsubsection*{Editing or Deleting Log Files} Since most NoSQL audit logs lack cryptographic integrity checks and are often stored as human-readable text files on the filesystem, a privileged user can directly modify or erase log files to remove evidence. Unlike relational systems that may store audit data in protected internal tables (e.g., Oracle's \texttt{sys.aud\$}), file-based logs are vulnerable to standard shell commands. An attacker with shell access could truncate a log file with \texttt{> /var/log/app.log} or use a text utility like \texttt{sed} to surgically remove specific log entries matching a pattern, such as \texttt{sed -i '/DELETE.*secret\_data/d' /var/log/app.log}.

\subsubsection*{Bypassing Application-Layer Controls} Actions taken at the deployment layer occur outside the purview of the database engine itself. A DO can use environment variables like \texttt{LD\_PRELOAD} to inject a malicious shared library that intercepts and filters logging calls before they reach the application's code. Because this is a change to the execution environment, not the application code, it effectively bypasses any internal controls and does not trigger application-level alerts.

\fi
\section{Experiments}
\label{sec:experiments}
Our experiments evaluate the framework across both key–value and document-oriented NoSQL databases to capture a wide range of underlying storage architectures. 
The selected engines collectively represent contrasting design paradigms, including in-place modification, append-only logging, and copy-on-write versioning. 
This diversity ensures that the framework’s recovery and correlation capabilities are validated under multiple operational models and data layouts. 
We conducted all experiments on both Windows and Linux systems to verify portability. 
The experimental environment for the performance metrics reported in this section consisted of a workstation running Ubuntu~22.04~LTS, equipped with an AMD~Ryzen~9~7950X3D processor (16~cores, 32~threads @~4.2~GHz) and 64~GB of RAM.

\paragraph{Dataset}
\label{sec:dataset}

We used the Star Schema Benchmark (SSBM)\cite{SSBM} to generate the experimental dataset. The relational structure of the SSBM data was adapted to fit the two NoSQL models used in our experiments: document and key-value. For the document database, we converted each row from an SSBM table into a single JSON document, where the original column names became the fields of the document. For the key-value databases, we transformed each row into a single key-value pair by selecting two representative columns to serve as the key and value, as detailed in Table~\ref{tab:ssbm_to_kv_mapping}.

\begin{table}[h]
\centering
\begin{tabular}{|l|l|l|}
\hline
\textbf{SSBM Table} & \textbf{Key Column} & \textbf{Value Column} \\ \hline
Customer            & name                & nation                \\ \hline
Supplier            & name                & city                  \\ \hline
Part                & partkey             & name                  \\ \hline
Date                & datekey             & date                  \\ \hline
Lineorder           & orderkey            & shipmode              \\ \hline
\end{tabular}
\caption{SSBM table mapping for LMDB and ZODB.}
\label{tab:ssbm_to_kv_mapping}
\end{table}

To create a unique and identifiable key for each record across all models, we standardized the key-generation process. For tables with integer primary keys such as \texttt{Part}, \texttt{Date}, and \texttt{Lineorder}, we prefixed the table name to the zero-padded integer key (e.g., \texttt{"Part\#00000001"}). For tables like \texttt{Customer} and \texttt{Supplier}, the existing name field served as the key. This consistent keying scheme allowed for uniform referencing of records regardless of the underlying database model.

\paragraph{Acquisition} All snapshots in this section were captured by the SA under the host-level policy in Section~\ref{sec:threat-model} (Figure~\ref{fig:file_write_event_detection}, Table~\ref{table:snapshot_policy}). Events classified as \emph{Normal} serve as baselines and are logged only; \emph{Maintenance} operations produce informational snapshots; and \emph{Suspicious} events (e.g., non-service writers, audit suppression, binary-integrity mismatches, or drift anomalies) trigger immediate forensic snapshots. Accordingly, by default we analyze a single \emph{After} snapshot captured automatically upon the first \emph{Suspicious} event. A \emph{Baseline} snapshot (captured as an informational snapshot under an approved \emph{Maintenance} window) is included only for experiments that require comparative analysis, specifically the in-place modification and the deletion case.

\if false
\subsection{Experiment 1: Execution \& Performance Measurement}
\label{sec:exp_performance}

The goal of this experiment is to measure the execution performance of the \radar{} framework across different NoSQL storage architectures. To create a realistic forensic scenario, we first populated each database with one million records and then executed a mixed workload of several thousand logged and unlogged inserts, updates, and deletes to generate a comprehensive application audit log. The evaluation then measures the end-to-end processing time, which encompasses two distinct stages reflective of \radar{}'s architecture (Figure \ref{fig:radar-workflow}): first, the initial carving of raw database files by the Automated NoSQL Carver (ANOC), and second, the subsequent analysis where \radar{}'s correlation engine ingests the carved artifacts and the audit log to identify unattributed operations. We conducted this experiment on three representative databases, each with a distinct storage model: LiteDB (in-place modification), LMDB (Copy-on-Write), and ZODB (append-only object store). The performance metrics reported represent the average of several trial runs to ensure consistency and mitigate the effects of transient system load.

The first stage of execution involved carving the raw database files with ANOC. For LiteDB, whose in-place modification architecture necessitates a comparative analysis, two snapshots were required. The ``before'' state was captured as a 337 MB file. Following the workload, the ``after'' state was captured, resulting in a slightly larger 340 MB file due to the mixed operations. ANOC carved the first file at a rate of 15.4 MB/s, taking 21.88 seconds, and the second file at a rate of 15.2 MB/s, taking 22.37 seconds. This resulted in a total ANOC carving time of 44.25 seconds for LiteDB. For LMDB and ZODB, which support single-snapshot analysis, only one file was processed. ANOC carved the 100 MB LMDB file at 29.2 MB/s in 3.42 seconds and the 121 MB ZODB file at 18.70 MB/s in 6.47 seconds.

In the second stage, \radar{} processed the carved data alongside corresponding audit logs. For LiteDB, ANOC produced 668 MB of carved artifacts from the ``before'' snapshot and 677 MB from the ``after'' snapshot. \radar{} then analyzed the combined 1.3 GB of carved data against a 715 MB audit log. Based on the total 2.01 GB of data processed, it achieved a high throughput of 80.2 MB/s, completing the analysis in 25.69 seconds. For LMDB, it processed 212 MB of carved data and a 187 MB log (399 MB total) at a rate of 23.7 MB/s, taking 16.83 seconds. Similarly, for ZODB, the analysis of 235 MB of carved data and a 190 MB log (425 MB total) occurred at 21.3 MB/s, taking 19.92 seconds. The total end-to-end execution time for the \radar{} framework to find unattributed operations was 69.94~seconds for LiteDB, 20.25~seconds for LMDB, and 26.39~seconds for ZODB. Table \ref{table:exp1-summary} summarizes the stage-wise and total timings for quick comparison across engines.

\fi

\subsection{Experiment 1: Execution Time Measurement}
\label{sec:exp_performance}


\begin{table}[!t]
\renewcommand{\arraystretch}{1.2}
\setlength{\tabcolsep}{5pt}
\centering
\small
\begin{tabular}{|l|l|r|r|r|r|r|}
\hline
\multirow{2}{*}{\textbf{DB}} & \multicolumn{1}{c|}{\textbf{Snapshot}} & \multicolumn{1}{c|}{\textbf{\NoCarver{}}} & \multicolumn{1}{c|}{\textbf{Artifacts}} & \multicolumn{1}{c|}{\textbf{Log File}} & \multicolumn{1}{c|}{\textbf{\radar{}}} & \multicolumn{1}{c|}{\textbf{Total}} \\
& \multicolumn{1}{c|}{\textbf{(MB)}} & \multicolumn{1}{c|}{\textbf{(m)}} & \multicolumn{1}{c|}{\textbf{(MB)}} & \multicolumn{1}{c|}{\textbf{(MB)}} & \multicolumn{1}{c|}{\textbf{(m)}} & \multicolumn{1}{c|}{\textbf{(m)}} \\
\hline
LMDB   & 1 file (100)       & 0.11  & 212 & 187 & 2.25 & 2.36 \\
\hline
ZODB   & 1 file (121)       & 0.24  & 235       & 190 & 2.42 & 2.66 \\
\hline
BerkeleyDB & 2 files (62, 64) & 0.12 + 0.14 & 105 + 108 & 184 & 0.71 & 1 \\
\hline
MDBX & 1 file (70) & 0.15 & 126 & 180 & 1.29 & 1.44 \\
\hline
etcd & 1 file (135) & 0.41 & 239 & 186 & 2.48 & 2.89 \\
\hline
Durus & 1 file (120)  & 0.21 & 231 & 189 & 2.36 & 2.57 \\
\hline
LiteDB & 2 files (603, 607) & 30.71 + 31.33 & 1751 + 1752 & 712 & 2.21 & 64.25 \\
\hline
RavenDB & 1 file (512) & 32.16 & 1755 & 711 & 41.33 & 73.49\\
\hline
Realm & 1 file (159) & 33.41 & 1757 & 713 & 42.17 & 75.58\\
\hline
NitriteDB & 1 file (724) & 34.93 & 1758 & 712 & 42.89 & 77.82\\
\hline
\end{tabular}
\caption{Experiment 1 summary (sizes in MB, times in minutes)}
\label{table:exp1-summary}
\end{table}

The goal of this experiment is to measure the total execution time of \radar{} framework across \emph{ten} NoSQL databases spanning different storage architectures (e.g., Copy-on-Write, append-only, and in-place modification). We first populated each database with one million records and then executed a mixed workload of several thousand logged and unlogged inserts, updates, and deletes to produce residual artifacts and a corresponding application audit log. The total processing time is computed as the sum of two stages aligned with \radar{}'s workflow (Figure~\ref{fig:radar-workflow}): (i) ANOC carving of the raw database snapshot(s), and (ii) \radar{} correlation over the carved artifacts and the audit log to identify unattributed operations.

Table~\ref{table:exp1-summary} summarizes snapshot sizes (MB), ANOC carving time (minutes), carved artifact sizes (MB), log sizes (MB), \radar{} correlation time (minutes), and the combined total time (minutes). For in-place engines (LiteDB and BerkeleyDB), we collected ``before'' and ``after'' snapshots. When a snapshot comprises multiple files, ANOC processes each file and the carving time and artifact sizes are summed for the total. Most engines with smaller snapshots (LMDB, ZODB, BerkeleyDB, MDBX, etcd, and Durus) complete in under 3 minutes. In contrast, the largest snapshots drive the longest runtimes. LiteDB’s total is dominated by two-snapshot carving (64.25 minutes total), while RavenDB, Realm, and NitriteDB are dominated by correlation over $\sim$1750~MB of carved artifacts plus $\sim$711~MB of logs (73.49--77.82 minutes total).

LiteDB and BerkeleyDB achieve lower \radar{} correlation time despite requiring two snapshots because \radar{} first performs a fast MD5 page-hash comparison and only performs record-level analysis on pages with hash mismatches. For engines analyzed using a single snapshot (e.g., Copy-on-Write or append-only stores), \radar{} must reconcile the full set of carved (often historical) artifacts against the audit log, which is inherently more expensive.

\if false
\begin{table}[!t]
\renewcommand{\arraystretch}{1.2}
\setlength{\tabcolsep}{5pt}
\centering
\small
\begin{tabular}{|l|l|r|r|r|r|r|}
\hline
\multirow{2}{*}{\textbf{DB}} & \multicolumn{1}{c|}{\textbf{Snapshot}} & \multicolumn{1}{c|}{\textbf{\NoCarver{}}} & \multicolumn{1}{c|}{\textbf{Artifacts}} & \multicolumn{1}{c|}{\textbf{Log File}} & \multicolumn{1}{c|}{\textbf{\radar{}}} & \multicolumn{1}{c|}{\textbf{Total}} \\ 
& \multicolumn{1}{c|}{\textbf{(MB)}} & \multicolumn{1}{c|}{\textbf{(s)}} & \multicolumn{1}{c|}{\textbf{(MB)}} & \multicolumn{1}{c|}{\textbf{(MB)}} & \multicolumn{1}{c|}{\textbf{(s)}} & \multicolumn{1}{c|}{\textbf{(s)}} \\ 
\hline
LiteDB & 2 files (337, 340) & 44.25 & 668 + 677 & 715 & 25.69 & 69.94 \\
\hline
LMDB   & 1 file (100)       & 3.42  & 212       & 187 & 16.83 & 20.25 \\
\hline
ZODB   & 1 file (121)       & 6.47  & 235       & 190 & 19.92 & 26.39 \\
\hline
\end{tabular}
\caption{Experiment 1 summary (sizes in MB, times in seconds)}
\label{table:exp1-summary}
\end{table}
\fi


\if false

\subsection{Experiment 2: Detecting Deletion in Append-Only DBMS}
\label{sec:exp2-deleted}

This experiment evaluates \radar{}’s ability to detect deleted records that remain physically present in storage but have no corresponding entry in the audit log. Such discrepancies indicate unlogged or tampered deletions (\emph{Unattributed Deletes}). We examined two append-only databases, \textit{ZODB} and \textit{NitriteDB}, under identical workloads to assess how their storage behavior exposes residual data for forensic recovery.

\paragraph{Procedure.}

We inserted 30{,}000 records derived from the SSBM Customer table (Section~\ref{sec:dataset}).
In ZODB, each entry used the customer name as the key and the nation as the value. In NitriteDB, each customer record existed as a single document. We then issued two deletion operations:
\begin{itemize}
\item \textbf{Delete 1 (legitimate):} Deleted keys \texttt{Customer\#000000100}–\texttt{Customer\#000000120} (\textbf{21 records}) with logging enabled.
\item \textbf{Delete 2 (malicious):} Deleted keys \texttt{Customer\#000000121}–\texttt{Customer\#000000130} (\textbf{10 records}) with logging disabled.
\end{itemize}

After performing \texttt{Delete 2} (malicious), we captured file snapshots from both databases and supplied them to \NoCarver.
\NoCarver{} analyzed each snapshot to reconstruct all logical records directly from storage. It sorted each key’s sequence of revisions by physical offset and transaction order: newer revisions appeared at higher offsets, while older ones persisted at lower offsets.
We labeled records that did not appear in the latest committed state but contained residual bytes from prior transactions as \emph{Deleted}.
These carved \emph{Deleted} states formed the set $C$. \radar{} compared $C$ against the audit log $L$ using Algorithm~\ref{alg:detect-deletes}.
When \radar{} found a record whose exact \texttt{(key,value)} pair lacked a corresponding \texttt{delete} or \texttt{update(old\_value)} entry in $L$, it added that record to $\Delta_{\mathrm{del}}$ as an \emph{Unattributed Delete}.

\paragraph{ZODB Results.}
In ZODB, each transaction appends a new object revision sequentially and leaves prior bytes intact.
ANOC recovered both active and obsolete revisions, with older versions appearing at lower offsets.
During carving, ANOC identified records missing from the latest committed transaction but still present as earlier revisions and marked them as \emph{Deleted}.
\radar{} matched all 21 legitimate deletions (\emph{Delete 1}) to their corresponding \texttt{delete(key,value)} entries in the log. It flagged the 10 malicious deletions (\emph{Delete 2}) as \textbf{Unattributed Deletes} because no log entries existed for them.

\paragraph{NitriteDB Results.}
NitriteDB’s underlying MVStore appends each commit as a new version of the storage file and retains earlier chunks until compaction.
ANOC traced the version lineage of each record and located older documents in previous store versions that no longer appeared in the latest map.
It labeled these earlier documents as \emph{Deleted}.
\radar{}’s audit reconciliation identified the same ten unlogged deletions (\emph{Delete 2}) as \textbf{Unattributed Deletes} and confirmed that the legitimate range (\emph{Delete 1}) aligned with logged events.

Both append-only databases exhibited consistent physical evidence of deletion.
In ZODB, older revisions persisted at lower offsets, while in NitriteDB, they remained within earlier store versions even after logical removal.
ANOC carved these residual historical states, and \radar{}’s audit correlation algorithm (Algorithm~\ref{alg:detect-deletes}) isolated those without matching log entries.
These findings show that a single snapshot taken before packing or compaction can reveal unlogged deletions through the combined analysis of storage-level continuity and audit-log divergence.


\subsection{Experiment 3: Post-Compaction Insert/Update Detection}
\label{sec:exp3-postcompaction}

This experiment evaluates \radar{}’s ability to detect unlogged inserts and updates after an adversary uses a maintenance operation to remove historical bytes from storage. Such behavior conceals prior versions by rewriting the database so that only the most recent values remain. Two append-oriented databases, \textit{ZODB} and \textit{NitriteDB}, were tested under identical workloads to determine how packing and compaction affect the visibility of unlogged changes.

\paragraph{Procedure.}

We used the \textbf{Customer} table containing 30{,}000 records. Two unlogged modifications were issued, followed by a maintenance routine intended to eliminate historical content:
\begin{itemize}
  \item \textbf{Action 1 (Unlogged Update):} Modified \texttt{Customer\#000000200} from \\ \texttt{NATION="GERMANY"} to \texttt{NATION="MALICIOUS\_UPDATE"}.
  \item \textbf{Action 2 (Unlogged Insert):} Added a new record \texttt{Customer\#999999} with \\ \texttt{NATION="ROGUE\_AGENT"}.
  \item \textbf{Action 3 (Cover Tracks — Maintenance):}
    \begin{itemize}
      \item In ZODB, executed the \texttt{pack} utility to rewrite the database and retain only objects reachable in the most recent transaction.
      \item In NitriteDB, triggered MVStore compaction, which rewrote all live data into new chunks and pruned older ones.
    \end{itemize}
\end{itemize}

After maintenance, we captured a single suspicious snapshot and analyzed it using \NoCarver.
Because both engines removed obsolete revisions, the carved dataset contained only active records corresponding to the final state.
\radar{} then compared these carved records against the audit log using Algorithm~\ref{alg:detect-inserts}, which identified \emph{unattributed inserts}, records whose \texttt{(key,value)} pairs were present on disk but not explained by any \texttt{insert} or \texttt{update(new\_value)} in the log.
It then applied Algorithm~\ref{alg:detect-updates} to link any corresponding deleted and active versions by key and detect \emph{unattributed updates} based on mismatched values.

\paragraph{ZODB Results.}
After packing, ZODB retained only the latest version of each record.  
The carved snapshot included \texttt{Customer\#999999} and \texttt{Customer\#000000200} with the value \texttt{"MALICIOUS\_UPDATE"}.  
\radar{}’s insert-detection phase (Algorithm~\ref{alg:detect-inserts}) marked both as \emph{Unattributed Inserts}, as no matching creation or update entries existed in the audit log.  
When Algorithm~\ref{alg:detect-updates} compared against deleted-state records, no prior values remained, packing had removed them entirely, so no update pairs could be formed.  
The altered record therefore appeared indistinguishable from a fresh, unlogged insertion.  
This confirmed that ZODB’s pack operation effectively erases prior transaction bytes, leaving only the final state for carving.

\paragraph{NitriteDB Results.}
In NitriteDB, MVStore compaction yielded a similar outcome but within a document-oriented context.
Each record was stored as a JSON document representing a customer entry, including attributes such as \texttt{"c\_custkey"}, \texttt{"c\_name"}, and \texttt{"c\_nation"}, among others.
After compaction, only the most recent documents survived; older chunks containing prior \texttt{"c\_nation"} values were dropped.
ANOC recovered the active documents for \texttt{Customer\#999999} and \texttt{Customer\#000000200}, the latter reflecting a change in the \texttt{"c\_nation"} field from \texttt{"GERMANY"} to \texttt{"MALICIOUS\_UPDATE"}.
During \radar{}’s processing, Algorithm~\ref{alg:detect-inserts} again flagged both as \emph{Unattributed Inserts}, since neither had a corresponding \texttt{insert} or \texttt{update(new\_value)} event in the log.  
Algorithm~\ref{alg:detect-updates} attempted to locate deleted documents for the same keys but found none, confirming that compaction had eliminated all prior chunk-level states.

Both databases showed the same post-maintenance pattern.
Once packing or compaction eliminated historical bytes, unlogged updates became indistinguishable from new inserts.
ANOC carved only the surviving active states, while \radar{} applied its detection pipeline, first isolating unattributed inserts (Algorithm~\ref{alg:detect-inserts}), then consolidating possible update pairs (Algorithm~\ref{alg:detect-updates}).
The results demonstrate that a single post-maintenance snapshot can still reveal concealed insertions and value changes through audit-log divergence, even when underlying revision history has been physically removed.

\fi


\subsection{Experiment 2: Detecting Unattributed Operations in Append-Only Architectures}
\label{sec:exp2-appendonly}

The goal of this experiment is to evaluate \radar{}'s effectiveness in detecting unlogged deletions and modifications within append-only architectures (\textit{ZODB} and \textit{NitriteDB}) under two distinct storage scenarios.
In \textbf{Scenario A}, the database operates normally by appending new record revisions to the end of the file (no packing or compaction), allowing prior record states to persist as residual artifacts at earlier file offsets.
In \textbf{Scenario B}, an adversary attempts to conceal tampering by invoking a maintenance operation (ZODB \texttt{pack} or NitriteDB MVStore compaction) that rewrites the database and prunes these historical bytes.
Table~\ref{tab:append-only-results} provides an overview of these scenarios and the resulting detection outcomes. As shown, \radar{} adapts its detection logic by extracting deleted artifacts when history is preserved (Scenario~A) and falling back to identifying unexplained active records as unattributed inserts when history is erased (Scenario B).

\begin{table*}[ht]
\centering

\resizebox{\textwidth}{!}{%
\begin{tabular}{|l|l|l|l|l|}
\hline
\textbf{Scenario} & \textbf{Operation} & \textbf{Logging} & \textbf{\NoCarver{} Result} & \textbf{\radar{} Result} \\ \hline
\multirow{2}{*}{\begin{tabular}[c]{@{}l@{}}\textbf{A. Deletion}\\ (History Preserved)\end{tabular}}
& \begin{tabular}[c]{@{}l@{}}Delete 1 (Legitimate)\end{tabular}
& Enabled
& \begin{tabular}[c]{@{}l@{}}Active + Deleted\\(Historical)\end{tabular}
& Matched Log \\ \cline{2-5}
& \begin{tabular}[c]{@{}l@{}}Delete 2 (Malicious)\end{tabular}
& Disabled
& \begin{tabular}[c]{@{}l@{}}Active + Deleted\\(Historical)\end{tabular}
& 10 Unattributed Deletes \\ \hline
\multirow{2}{*}{\begin{tabular}[c]{@{}l@{}}\textbf{B. Maintenance}\\ (History Erased)\end{tabular}}
& \begin{tabular}[c]{@{}l@{}}Update (Malicious)\end{tabular}
& Disabled
& \begin{tabular}[c]{@{}l@{}}Active Only\\(History Pruned)\end{tabular}
& \begin{tabular}[c]{@{}l@{}}Unattributed Insert\\(Update Detected as Insert)\end{tabular} \\ \cline{2-5}
& \begin{tabular}[c]{@{}l@{}}Insert (Malicious)\end{tabular}
& Disabled
& \begin{tabular}[c]{@{}l@{}}Active Only\\(History Pruned)\end{tabular}
& Unattributed Insert \\ \hline
\end{tabular}%
}
\caption{Results summary for detecting unattributed operations in ZODB and NitriteDB.}
\label{tab:append-only-results}
\end{table*}

\paragraph{Procedure}
We used the SSBM \textbf{Customer} table and populated each database with 30{,}000 initial records.
In ZODB, each entry used the customer name as the key and the nation as the value.
In NitriteDB, each customer record was stored as a document.
We executed the two scenarios as independent runs, each starting from a fresh baseline database state, and captured a single forensic snapshot at the end of each run.
For each snapshot, we supplied the database files to \NoCarver{} to reconstruct logical records directly from storage.
\radar{} then reconciled the carved artifacts against the corresponding application audit log.

\begin{itemize}
    \item \textbf{Scenario A (History Preserved: Unlogged Deletions).}
    \begin{itemize}
        \item \textbf{Delete 1 (Legitimate):} Deleted keys \texttt{Customer\#000000100}--\texttt{Customer\#000000120} (21 records) with logging enabled.
        \item \textbf{Delete 2 (Malicious):} Deleted keys \texttt{Customer\#000000121}--\texttt{Customer\#000000130} (10 records) with logging disabled.
        \item \textbf{Snapshot:} Captured a snapshot immediately after Malicious Delete. No packing or compaction was performed.
    \end{itemize}

    \item \textbf{Scenario B (History Erased: Post-Maintenance Tampering).}
    \begin{itemize}
        \item \textbf{Update (Malicious):} Modified \texttt{Customer\#000000200} from \texttt{NATION="GERMANY"} to \texttt{NATION="MALICIOUS\_UPDATE"} with logging disabled.
        \item \textbf{Insert (Malicious):} Inserted \texttt{Customer\#999999} with \texttt{NATION="ROGUE\_AGENT"} with logging disabled.
        \item \textbf{Maintenance:} Executed ZODB \texttt{pack} and MVStore compaction in NitriteDB to rewrite live data and prune historical chunks.
        \item \textbf{Snapshot:} Captured a snapshot immediately after maintenance completed.
    \end{itemize}
\end{itemize}

\paragraph{Results: Scenario A (Detection with History)}
In the absence of maintenance operations, both databases preserved historical bytes that provided physical evidence of deletion.
For ZODB, each transaction appended new object revisions sequentially, leaving prior revisions intact at lower file offsets.
For NitriteDB, MVStore retained earlier versions and chunks.
Consequently, \NoCarver{} recovered both the active records and the deleted-state remnants for the targeted keys.
It labeled a record state as \emph{Deleted} when it remained physically present as a prior revision in storage but did not appear in the latest committed database state.
\radar{} utilized Algorithm~\ref{alg:detect-deletes} to compare these carved \emph{Deleted} artifacts against the audit log.
It matched all 21 legitimate deletions (Delete 1) to their logged \texttt{delete(key,value)} events and flagged the 10 malicious deletions (Delete 2) as \textbf{Unattributed Deletes} because no log entries existed for those deleted pairs.
These results confirm that when historical bytes persist, \radar{} can recover exact deleted values and identify unauthorized removals through audit-log divergence.

\paragraph{Results: Scenario B (Detection after Historical Erasure)}
The execution of ZODB \texttt{pack} and MVStore compaction removed the physical history. Consequently, \NoCarver{} recovered only the active records corresponding to the final state, with no deleted-state artifacts remaining to form update pairs.
Despite this loss of history, \radar{} detected the unlogged changes by comparing the carved active state to the audit log.
Using Algorithm~\ref{alg:detect-inserts}, \radar{} flagged both \texttt{Customer\#000000200} and \texttt{Customer\#999999} as \textbf{Unattributed Inserts}, since neither had a corresponding \texttt{update(new\_value)} or \texttt{insert} explanation in the log.
Algorithm~\ref{alg:detect-updates} could not form deleted--active pairs for \texttt{Customer\#000000200} because packing/compaction eliminated the prior versions.

\if false

\subsection{Experiment 4: Deletion Detection in Page-Based DBMS}
\label{sec:exp4-deletedpages}

This experiment evaluates \radar{}’s ability to detect unlogged deletions in page-based, Copy-on-Write (CoW) key–value stores. 
We used \textit{LMDB} and \textit{MDBX}, two closely related engines that organize data within a CoW B\textsuperscript{+}-tree structure. 
When a record is deleted, the database rewrites the affected leaf page to a new location and updates the parent internal nodes to reference it. 
The previous page remains physically intact, marked as stale, and later becomes eligible for reuse. 
These transient remnants provide valuable opportunities for forensic recovery.

\paragraph{Procedure.}

We used the \textbf{Supplier} table containing 20{,}000 records derived from the SSBM dataset (Section~\ref{sec:dataset}). 
We issued two deletion operations to emulate legitimate and malicious activity:
\begin{itemize}
  \item \textbf{Delete 1 (legitimate):} Deleted keys \texttt{Supplier\#000000010}–\texttt{Supplier\#000000030} with audit logging enabled.
  \item \textbf{Delete 2 (malicious):} Deleted keys \texttt{Supplier\#000000110}–\texttt{Supplier\#000000120} with logging disabled.
\end{itemize}

After performing \texttt{Delete 2} (malicious), we captured a snapshot of each database file and analyzed it using \NoCarver.

ANOC first reached the most recent meta page to obtain the current database root, then traversed the B\textsuperscript{+}-tree from the root through branch pages to enumerate all reachable leaf pages and marked their records as \emph{Active}. 
It then scanned the file for leaf pages that were not reachable from the current tree and labeled the records on those orphaned pages as \emph{Deleted}. ANOC’s carved output included key–value records with their page identifiers and status metadata, which \radar{} subsequently correlated with the audit log.

\paragraph{LMDB Results.}
In LMDB, ANOC identified several leaf pages no longer referenced by the active B\textsuperscript{+}-tree yet still containing complete \texttt{Supplier} records.  
These pages corresponded to the ranges affected by \texttt{Delete 1} and \texttt{Delete 2}.  
\radar{} compared each carved record against the audit log using Algorithm~\ref{alg:detect-deletes}.  
It successfully matched all records from \texttt{Delete 1} to logged \texttt{delete(key,value)} operations.  
The ten records from \texttt{Delete 2} lacked corresponding entries, and \radar{} flagged them as \textbf{Unattributed Deletes}.  
Subsequent writes caused LMDB to reuse the affected pages, confirming that its page reclamation process removes residual evidence once it reallocates freed space.

\paragraph{MDBX Results.}
MDBX displayed nearly identical behavior due to its shared lineage with LMDB.  
ANOC reconstructed both the active and stale page sets by traversing the meta area and identifying orphaned leaf nodes.  
The deleted ranges again produced stale pages containing recoverable \texttt{Supplier} records.  
\radar{} applied the same comparison process as with LMDB: it attributed legitimate deletions to audit-log entries and flagged the unlogged range (\texttt{Delete 2}) as ten \textbf{Unattributed Deletes}.  
Because MDBX performs more aggressive page recycling during transactions, it overwrote some deleted pages earlier than LMDB.  
Despite this difference, both engines exhibited the same forensic pattern, transient visibility of deleted data followed by physical reuse.

Both databases revealed deleted-state artifacts residing on unreferenced pages. 
ANOC accurately distinguished between active and stale pages, and \radar{} confirmed which records lacked corresponding audit events.

\subsection{Experiment 5: Unattributed Update Detection in CoW Page-Based DBMS}
\label{sec:exp5-updates}

This experiment evaluates \radar{}’s ability to detect unlogged updates within page-based databases that perform Copy-on-Write (CoW) rewriting during modifications. 
We used \textit{LMDB} and \textit{MDBX}, both of which rewrite leaf pages on update and temporarily retain prior versions until those pages are reused or compacted.

\paragraph{Procedure.}

We used the \texttt{Supplier} table containing 20{,}000 records (Section~\ref{sec:dataset}). 
We executed two update ranges:

\begin{itemize}
  \item \textbf{Update 1 (legitimate):} Updated keys \texttt{Supplier\#000000200}–\texttt{Supplier\#000000220} to new values with audit log enabled.
  \item \textbf{Update 2 (malicious):} Updated keys \texttt{Supplier\#000000221}–\texttt{Supplier\#000000230} to new values with audit log disabled.
\end{itemize}

We captured snapshots at four timepoints to observe how pre-update pages evolved under different storage conditions:
\begin{enumerate}
  \item \textbf{$T_0$ Baseline.} Captured a pre-update snapshot to verify the absence of stale artifacts.
  \item \textbf{$T_1$ Post-update (reader pinned).} Opened a long-lived read transaction before issuing the updates and kept it open through $T_1$ to pin pre-update pages. Captured a snapshot after committing both updates.
  \item \textbf{$T_2$ Reuse stressor.} Closed the read transaction, inserted \textbf{5{,}000} additional records e.g., \texttt{Supplier\#000020001}–\texttt{Supplier\#000025000}, and captured another snapshot to induce freelist reuse and overwrite stale pages.
\end{enumerate}

At each timepoint, we analyzed the snapshot using \NoCarver. 
ANOC reached the current meta root and traversed the B\textsuperscript{+}-tree to enumerate active leaf pages, marking their records as \emph{Active}. 
It then carved leaf pages that the active tree no longer referenced and labeled their records as \emph{Deleted}. 
\radar{} applied Algorithm~\ref{alg:detect-inserts} to surface active records lacking corresponding \texttt{insert} or \texttt{update(new\_value)} entries and Algorithm~\ref{alg:detect-updates} to pair keys appearing across the deleted and active sets with differing values. 
Each valid pair—an older \emph{Deleted} record and its newer \emph{Active} version—was classified as an \emph{Unattributed Update}. 
Legitimate updates reconciled through matching \texttt{update(new\_value)} log entries.

\paragraph{LMDB Results.}
At $T_1$, ANOC recovered both sides of each change for the modified keys: pre-update values on stale pages (carved as \emph{Deleted}) and post-update values on active pages (marked \emph{Active}). 
\radar{} matched all \textbf{21} keys from \textbf{Update 1} to their audit-log entries. 
For \textbf{Update 2}, all \textbf{10} keys formed valid \emph{Deleted}–\emph{Active} pairs without corresponding log events, and \radar{} flagged them as \textbf{Unattributed Updates}. 
After the write burst at $T_2$ (5{,}000 inserts), LMDB reused the stale pages; the pre-update side of every pair disappeared, leaving only the active values. 
\radar{} therefore classified all \textbf{10/10} Update~2 keys as \emph{Unattributed Inserts}.

\paragraph{MDBX Results.}
MDBX exhibited the same CoW signature at $T_1$. 
ANOC carved pre-update values from orphaned pages and identified post-update values on active pages. 
\radar{} reconciled all legitimate updates from \textbf{Update 1} and flagged the ten keys from \textbf{Update 2} as \textbf{Unattributed Updates}. 
Following the reuse workload at $T_2$ (5{,}000 inserts), MDBX reclaimed stale pages aggressively; no pre-update remnants remained, and all affected keys appeared as \emph{Unattributed Inserts}.

Both engines displayed a dual-state footprint after unlogged updates ($T_1$): an active page storing the new value and an orphaned page retaining the prior version. 
ANOC separated these states by reachability, and \radar{} paired them by key to expose \emph{Unattributed Updates}. 
After induced page reuse at $T_2$, the deleted side vanished in both engines; however, \radar{} still surfaced the same events by classifying the remaining active records as \emph{Unattributed Inserts}. 
This shows that the \radar{} detects unlogged modifications both before and after page reuse.

\fi

\subsection{Experiment 3: Detecting Unattributed Operations in CoW Page-Based Architectures}
\label{sec:exp3-pagebased}

The goal of this experiment is to evaluate \radar{}'s capability to detect unlogged deletions and modifications in B\textsuperscript{+}-tree architectures (\textit{LMDB} and \textit{MDBX}) under two distinct storage scenarios governed by the Copy-on-Write (CoW) mechanism.
In \textbf{Scenario A}, the database handles updates and deletions by rewriting affected leaf pages to new locations while leaving prior versions physically intact as ``stale'' pages, enabling the direct recovery of deleted artifacts.
In \textbf{Scenario B}, we simulate an adversary attempting to conceal tampering by generating heavy write activity that induces page reuse, causing stale pages to be reclaimed and overwritten and thereby removing pre-update remnants.
Table~\ref{tab:page-based-results} summarizes how \radar{} detects tampering across these transient storage states. As shown, \radar{} flags unlogged deletions and updates by recovering stale-page artifacts when prior versions persist (Scenario~A and Snapshot $T_1$ in Scenario~B), and after page reuse overwrites those remnants (Snapshot $T_2$), it falls back to reporting the same tampering as \textbf{Unattributed Inserts} because update pairing is no longer possible.

\begin{table*}[ht]
\centering
\resizebox{\textwidth}{!}{%
\begin{tabular}{|l|l|l|l|l|}
\hline
\textbf{Scenario} & \textbf{Operation} & \textbf{Logging} & \textbf{\NoCarver{} Result} & \textbf{\radar{} Result} \\ \hline
\multirow{2}{*}{\begin{tabular}[c]{@{}l@{}}\textbf{A. Deletion}\\ (Stale Pages Intact)\end{tabular}}
& Delete 1 (Legitimate) & Enabled & \begin{tabular}[c]{@{}l@{}}Active + Deleted\\(Stale Pages)\end{tabular} & \begin{tabular}[c]{@{}l@{}}Authorized\\(Matched Log)\end{tabular} \\ \cline{2-5}
& Delete 2 (Malicious) & Disabled & \begin{tabular}[c]{@{}l@{}}Active + Deleted\\(Stale Pages)\end{tabular} & 10 Unattributed Deletes \\ \hline
\multirow{3}{*}{\begin{tabular}[c]{@{}l@{}}\textbf{B. Update \& Reuse}\\ (Pre/Post Reuse)\end{tabular}}
& Update 1 (Legitimate) & Enabled & \begin{tabular}[c]{@{}l@{}}Active + Deleted\\(Stale Pages)\end{tabular} & \begin{tabular}[c]{@{}l@{}}Authorized\\(Matched Log)\end{tabular} \\ \cline{2-5}
& Update 2 (Malicious) ($T_1$) & Disabled & \begin{tabular}[c]{@{}l@{}}Active + Deleted\\(Stale Pages)\end{tabular} & 10 Unattributed Updates \\ \cline{2-5}
& Insert (Malicious) ($T_2$) & Disabled & \begin{tabular}[c]{@{}l@{}}Active Only\\(Stale Overwritten)\end{tabular} & \begin{tabular}[c]{@{}l@{}}Unattributed Insert\\(Update Reported as Insert)\end{tabular} \\ \hline
\end{tabular}%
}
\caption{Results summary for detecting unattributed operations in LMDB and MDBX.}
\label{tab:page-based-results}
\end{table*}

\paragraph{Procedure}
We utilized the SSBM \textbf{Supplier} table, populating both databases with 20,000 initial records.
We then executed two scenarios:

\begin{itemize}
    \item \textbf{Scenario A (Stale Pages Intact: Unlogged Deletion).}
    \begin{itemize}
        \item \textbf{Delete 1 (Legitimate):} Deleted keys \texttt{Supplier\#000000010} to \texttt{Supplier\\\#000000030} with logging enabled.
        \item \textbf{Delete 2 (Malicious):} Deleted keys \texttt{Supplier\#000000110} to \texttt{Supplier\#000000119} with logging disabled.
        \item \textbf{Snapshot:} Captured immediately after Delete 2. No heavy write load was induced, ensuring stale pages remained unreclaimed.
    \end{itemize}

    \item \textbf{Scenario B (Pre/Post Reuse: Unlogged Operations).}
    \begin{itemize}
        \item \textbf{Update 1 (Legitimate):} Updated keys \texttt{Supplier\#000000200} to \texttt{Supplier \#000000220} with logging enabled.
        \item \textbf{Update 2 (Malicious):} Updated keys \texttt{Supplier\#000000221} to \texttt{Supplier\\ \#000000230} with logging disabled.
        \item \textbf{Snapshot $T_1$ (Pre-Reuse):} Opened a long-lived read transaction \emph{before} issuing Update 1 and 2 and kept it open to pin pre-update pages. This prevents those pages from being reclaimed for reuse, ensuring both versions remain observable at $T_1$. Captured a snapshot after both updates committed.
        \item \textbf{Insert (Force Reuse):} Closed the read transaction and inserted 5,000 new records (\texttt{Supplier\#000020001} to \texttt{Supplier\#000025000}) to induce freelist reuse and overwrite stale pages.
        \item \textbf{Snapshot $T_2$ (Post-Reuse):} Captured a snapshot after the insert burst completed.
    \end{itemize}
\end{itemize}

\paragraph{Results: Scenario A (Detection via Stale Pages)}
In the absence of heavy subsequent writes, both LMDB and MDBX retained leaf pages containing the deleted records.
Although these pages were no longer reachable from the active B\textsuperscript{+}-tree root, \NoCarver{} identified formatted leaf pages not traversed in the active set and labeled their records as \emph{Deleted}.
\radar{} utilized Algorithm~\ref{alg:detect-deletes} to compare these artifacts against the audit log.
It matched the legitimate deletions (Delete 1) to logged events and flagged the malicious deletions (Delete 2) as \textbf{Unattributed Deletes}.
This confirms that CoW architectures preserve deletion evidence in stale pages until storage pressure forces reclamation.

\paragraph{Results: Scenario B (Resilience to Page Reuse)}
At snapshot $T_1$, the databases exhibited a dual-state footprint: active pages stored the new values, while stale pages retained pre-update values.
\radar{} paired these versions using Algorithm~\ref{alg:detect-updates} and flagged the unlogged modifications as \textbf{Unattributed Updates}.
At snapshot $T_2$, the insertion workload induced freelist reuse and overwrote the stale pages holding the pre-update remnants for \texttt{Supplier\#000000221} to \texttt{Supplier\#000000230}.
\NoCarver{} recovered only the active state, and \radar{} flagged the same ten keys as \textbf{Unattributed Inserts} via Algorithm~\ref{alg:detect-inserts}, since no corresponding \texttt{insert} or \texttt{update(new\_value)} explanation existed in the log.
Without the deleted-side remnants at $T_2$, update pairing is no longer possible, so the same unlogged modifications are surfaced through the remaining active records.

\if false

\subsubsection*{Experiment 6: Unattributed In-Place Modification Detection}
The goal of this experiment is to detect subtle, unlogged modifications in a database that performs in-place updates, such as LiteDB. Unlike append-only (ZODB) or Copy-on-Write (LMDB) systems, LiteDB overwrites existing records directly within their allocated page. This means that the previous version of a record is destroyed at the storage level, making it impossible to recover from a single snapshot. To address this, we employ comparative analysis between ``before'' and ``after'' snapshots using page-level hashes.

We used the \texttt{parts} table in LiteDB containing 200{,}000 records. The following sequence of actions was performed:
\begin{itemize}
\item \textbf{Action 1 (Baseline Snapshot)}: Under an approved SA maintenance ticket, a reference baseline snapshot was captured by copying \texttt{Parts.db} to \texttt{Parts\_before.db}.

\item \textbf{Action 2 (Malicious In-Place Update)}: The adversary edits \texttt{part\#000000500} in place, changing \texttt{P\_Container} from \texttt{"JUMBO CASE"} to \texttt{"SM PKG"}. Because LiteDB overwrites bytes on the same page (no historical copy in this snapshot), the attacker also removes the matching audit entry so there is no logged explanation for this specific change.

\item \textbf{Action 3 (Legitimate Noise on the Same Page)}: To make the page change look routine at a coarse level, the adversary performs two ordinary, logged deletions on that same page (\texttt{part\#000000501} and \texttt{part\#000000502}). These legitimate events explain the page hash change, while the silent field edit in \texttt{part\#000000500} remains unattributed at the record level.

\item \textbf{Action 4 (After Snapshot \& Carving — Suspicious)}: Immediately after Actions~2–3, the SA policy classified the provenance gap as \emph{Suspicious} and captured a forensic snapshot (\texttt{Parts\_after.db}). \NoCarver was then run to produce \texttt{parts\_after.jsonl} with per-page MD5 hash.
\end{itemize}

\radar{} correctly identified \textbf{page 25} as modified by detecting a change in its MD5 hash between the Normal and Suspicious snapshots. Although the database engine also modified several other pages for index maintenance, \radar{} correctly disregarded this system-level noise to focus on the data page containing the suspicious activity. Within this page, \radar{} matched the deletions of \texttt{part\#000000501} and \texttt{part\#000000502} to their audit log entries, correctly attributing them as legitimate. However, its comparative analysis of \texttt{part\#000000500} revealed a change in the \texttt{P\_Container} field from ``JUMBO CASE'' to ``SM PKG'' without a corresponding update event in the audit log. Consequently, \radar{} flagged the record as an \textbf{Unattributed Update}, thereby identifying the malicious in-place modification.

\fi


\subsection{Experiment 4: In-Place Modification Detection}
\label{sec:exp6-inplace}
The goal of this experiment is to evaluate \radar{}'s ability to detect subtle, unlogged modifications in in-place storage engines (\textit{LiteDB} and \textit{Berkeley DB}).
Unlike append-only or Copy-on-Write designs, in-place engines overwrite bytes within an existing page, so pre-change content does not persist within a single snapshot.
\radar{} therefore compares a \emph{before} snapshot ($T_1$) and an \emph{after} snapshot ($T_2$), uses page-level hashing to localize changed pages, and performs record-level differencing to reconcile benign page activity with the audit log.
Table~\ref{tab:exp4_inplace_results} summarizes the experimental result. As shown, \radar{} isolates the modified page, reconciles the logged deletions, and flags the unlogged in-place edit as an \textbf{Unattributed Update}.

\begin{table*}[ht]
\centering
\resizebox{\textwidth}{!}{%
\begin{tabular}{|l|l|l|l|l|}
\hline
\textbf{Scenario} & \textbf{Operation} & \textbf{Logging} & \textbf{\NoCarver{} Result} & \textbf{\radar{} Result} \\ \hline

\multirow{4}{*}{\begin{tabular}[c]{@{}l@{}}\textbf{In-Place} \\ \textbf{Modification}\end{tabular}}
& \begin{tabular}[c]{@{}l@{}}Snapshot (Baseline) ($T_1$)\end{tabular}
& Enabled
& \multicolumn{1}{c|}{--}
& \multicolumn{1}{c|}{--} \\ \cline{2-5}

& \begin{tabular}[c]{@{}l@{}}Update (Malicious):\\ part\#000000500\end{tabular}
& Disabled
& \multicolumn{1}{c|}{--}
& \multicolumn{1}{c|}{--} \\ \cline{2-5}

& \begin{tabular}[c]{@{}l@{}}Delete (Legitimate):\\ part\#000000501--502 \\ (Same Page)\end{tabular}
& Enabled
& \multicolumn{1}{c|}{--}
& \multicolumn{1}{c|}{--} \\ \cline{2-5}

& \begin{tabular}[c]{@{}l@{}}Compare snapshots:\\ $T_1$ vs.\ $T_2$\end{tabular}
& \multicolumn{1}{c|}{--}
& \begin{tabular}[c]{@{}l@{}} All active records \\ from $T_1$ \& $T_2$\end{tabular}
& 1 Unattributed Update \\ \hline

\end{tabular}%
}
\caption{Results summary for detecting unattributed operations in LiteDB and Berkeley DB.}
\label{tab:exp4_inplace_results}
\end{table*}

\paragraph{Procedure}

We used the \textbf{Parts} dataset with 20{,}000 records. 
In LiteDB, each record was stored as a document; in Berkeley DB, each entry used \texttt{partkey} as the key and  \texttt{p\_name} as the value. The following sequence was executed:

\begin{itemize}
  \item \textbf{Snapshot $T_1$ (Baseline):} Captured the baseline snapshot ($T_1$) before performing any operations.
  \item \textbf{Update (Malicious):} Edited \texttt{part\#000000500} in place, changing \texttt{P\_Name} from \texttt{"azure spring"} to \texttt{"SM PKG"} without a corresponding audit log.
  \item \textbf{Delete (Legitimate):} Performed two ordinary, logged deletions on that same data page (\texttt{part\#000000501} and \texttt{part\#000000502}) to create benign page activity.
  \item \textbf{Snapshot $T_2$:} Captured an \texttt{after} snapshot, $T_2$ and ran \NoCarver{} to compute per-page hashes and reconstruct records for both snapshots.
\end{itemize}

\NoCarver{} computed an MD5 hash for every data page within each snapshot. 
\radar{} then accepted both the \texttt{before} and \texttt{after} snapshots as input, compared their respective page–hash maps to identify pages whose contents had changed, and enumerated the records located on those changed pages. 
From these differences, \radar{} derived record-level deltas representing what existed only in the baseline snapshot (\(\Delta^{-}\)) and what appeared only in the modified snapshot (\(\Delta^{+}\)), which it subsequently reconciled with the audit log.

\[
\begin{aligned}
\Delta^{-} &= \bigcup_{p \in \text{ChangedPages}} 
   \{ (k, v_{\text{before}}) \;\mid\; \text{record on page } p \text{ exists only in \texttt{before}} \}, \\[6pt]
\Delta^{+} &= \bigcup_{p \in \text{ChangedPages}} 
   \{ (k, v_{\text{after}}) \;\mid\; \text{record on page } p \text{ exists only in \texttt{after}} \}.
\end{aligned}
\]

\radar{} reconciled these deltas with the audit log: it first applied Algorithm~\ref{alg:detect-inserts} to surface \emph{unattributed inserts} (\(\Delta^{+}\) without matching log evidence), then applied Algorithm~\ref{alg:detect-updates} by pairing keys across \(\Delta^{-}\) and \(\Delta^{+}\) with differing values to identify \emph{unattributed updates}. 
Legitimate deletions and inserts on the same page reconciled to their corresponding log entries and were excluded.

\paragraph{LiteDB Results.}
Page-level hashing narrowed the search to a single modified data page (e.g., \textbf{page 25}). 
Within that page, \radar{} matched the logged deletions of \texttt{part\#000000501} and \texttt{part\#000000502} to the audit log and removed them from consideration. 
It then paired \((\texttt{part\#000000500}, \texttt{"azure spring"}) \in \Delta^{-}\) with \((\texttt{part\#000000500}, \\ \texttt{"SM PKG"}) \in \Delta^{+}\) and found no corresponding \texttt{update(new\_value)} event. 
Algorithm~\ref{alg:detect-updates} therefore classified \texttt{part\#000000500} as an \textbf{Unattributed Update}, revealing the malicious in-place edit. 
Index-maintenance pages also changed between snapshots but produced no record-level discrepancies and were ignored.

\paragraph{Berkeley DB Results.}
The same procedure isolated a modified data page containing \texttt{part\\\#000000500}. 
\radar{} reconciled the two logged deletions on that page and then compared the \texttt{before}/\texttt{after} values for \texttt{part\#000000500}. 
As in LiteDB, it observed the transition \texttt{"azure spring"} \(\rightarrow\) \texttt{"SM PKG"} with no matching audit update and flagged the key as an \textbf{Unattributed Update} via Algorithm~\ref{alg:detect-updates}. 
Because Berkeley DB applies in-place overwrites, no historical bytes were available in a single snapshot; the detection hinged on the two-snapshot differential and page scoping.

These results show that in-place engines, whether document-oriented (LiteDB) or key–value (Berkeley DB), do not retain pre-change bytes within a single snapshot. 
By hashing pages and differencing records across \emph{before}/\emph{after} snapshots, ANOC and \radar{} isolate the exact key and field that changed and attribute only the legitimate page activity to the audit log. 
The remaining per-key value transition, lacking a corresponding log event, is surfaced as an \emph{Unattributed Update}.

\section{Future Work}
While \radar{} demonstrates strong capabilities for detecting unlogged operations, several avenues for future research and development remain.

\paragraph*{Real-time Detection} Currently, \radar{} operates on forensic snapshots captured after suspicious events. Future work could explore continuous monitoring approaches that perform incremental analysis as database files change, enabling real-time threat detection and reducing the window of vulnerability.

\paragraph*{Encrypted Database Support} Many production databases employ encryption at rest, which presents challenges for direct storage carving. Investigating methods to extend \radar{} to encrypted environments, perhaps through integration with key management systems or analysis of encrypted page structures, would broaden its applicability.

\paragraph*{Distributed and Clustered Systems} Modern NoSQL deployments often span multiple nodes with complex replication and sharding strategies. Extending \radar{} to correlate artifacts across distributed systems and detect coordinated tampering attempts across cluster members represents an important research direction.

\paragraph*{Machine Learning Enhancement} The current rule-based reconciliation could be augmented with machine learning models trained to identify anomalous patterns in carved artifacts that may indicate sophisticated tampering attempts. This could help detect novel attack patterns that evade traditional signature-based detection.

\paragraph*{Performance Optimization} While our current implementation demonstrates practical performance, opportunities exist for optimization through parallel processing, incremental carving, and caching of previously analyzed pages. These improvements would be particularly valuable for very large databases.

\paragraph*{Integration with SIEM Platforms} Developing standardized interfaces to integrate \radar{} with existing Security Information and Event Management (SIEM) systems would facilitate deployment in production environments and enable correlation with other security signals.

\paragraph*{Extension to Cloud-Native Databases} As organizations increasingly adopt cloud-native NoSQL services, adapting \radar{} to work with cloud storage abstractions and serverless database architectures presents both technical challenges and opportunities.

Future work in these areas will extend \radar{}’s capabilities beyond post-incident analysis toward continuous, tamper-aware security monitoring. By unifying low-level storage observation with higher-level anomaly detection and system integration, \radar{} can evolve into a robust platform for trustworthy detection of unauthorized data manipulation across both on-premise and cloud environments. The increasing complexity of modern database infrastructures underscores the need to strengthen integrity verification and operational resilience against insider and system-level threats.

\section{Conclusion}

In this work, we presented \radar{}, a log-independent forensic framework that addresses the critical challenge of detecting unattributed operations in NoSQL databases. By unifying low-level storage carving with high-level audit log verification, \radar{} provides an approach for identifying log-evasion activity even when privileged attackers manipulate or suppress logging mechanisms.

Our framework makes several key contributions to the field of database forensics. First, we formalized the threat model of a malicious DevOps Engineer who can manipulate deployment configurations and audit logs but cannot suppress OS-level traces. Second, we developed two complementary detection methodologies tailored to different storage architectures: single-snapshot analysis for append-only and Copy-on-Write systems that preserve historical artifacts, and comparative analysis for in-place modification systems that overwrite prior states. Third, we introduced formal algorithms for identifying unattributed insertions, deletions, and updates through systematic reconciliation of carved artifacts with audit logs.

Through extensive experimentation across different representative NoSQL databases, including key-value and document stores, we demonstrated \radar{}'s effectiveness in detecting various tampering scenarios. Our results show that the framework successfully identifies unlogged operations even after maintenance activities such as packing or compaction that physically remove historical bytes. The framework achieved processing rates ranging from 31.7 to 397 MB/min depending on the database architecture, demonstrating practical performance for forensic investigations.

Importantly, \radar{}'s design philosophy of treating carved storage as ground truth rather than relying solely on potentially compromised logs represents a paradigm shift in database security monitoring. By operating at the intersection of application-level semantics and storage-level artifacts, the framework provides defense-in-depth against insider threats that traditional audit-based approaches cannot detect.

\section*{Acknowledgments} 
This work was partially funded by the Louisiana Board of Regents Grant LEQSF(2022-25)-RD-A-30.

\bibliographystyle{elsarticle-num} 
\bibliography{references} 

\end{document}